\documentclass[12pt,a4paper,oneside]{book}
\usepackage[T1]{fontenc}
\usepackage[latin2]{inputenc}
\usepackage{graphicx}
\usepackage{epsfig}
\def\ha{H_1}
\def\hb{H_2}
\def\DD{{\cal D}}
\newcommand{\bee}{\begin{equation}}
\newcommand{\ee}{\end{equation}}
\def\lsi{\raise0.3ex\hbox{$<$\kern-0.75em\raise-1.1ex\hbox{$\sim$}}}
\def\gsi{\raise0.3ex\hbox{$>$\kern-0.75em\raise-1.1ex\hbox{$\sim$}}}
\newcommand{\lsim}{\mathop{\lsi}}

\normalsize
\title{Astroparticle physics signals beyond the Standard Model}
\author{{\large S\'andor D. Katz}\\ PhD thesis\\
\\
advisor: Zolt\'an Fodor\\
\\
physics program\\
program leader: P\'eter Sz\'epfalusi\\
subprogram leader: George P\'ocsik\\
\\
\\
Institute for Theoretical Physics\\
E\"otv\"os University, Budapest\\
}
\begin{document}
\maketitle
\tableofcontents
\chapter{Introduction}\label{intro}
The Standard Model (SM) of particle physics is in perfect
agreement with all present accelerator experiments 
\footnote{Recently the anomalous magnetic moment of the muon has been
found to be
slightly different from the SM prediction \cite{muon}}. There are, however,
signals that point beyond the SM. These can be either theoretical
or experimental indications. On the theoretical side one could mention the
problems the SM suffers from: too many parameters, triviality of the Higgs
sector, Landau-pole in the U(1) sector, etc.
On the experimental side there are only indirect signals. In my thesis
I will deal with two such indications.

The first one is the presence of baryonic matter around us. The SM
seems to provide a mechanism for producing nonzero baryon number starting
from symmetric initial conditions. This requires a strong first order finite 
temperature
electroweak phase transition. 
However, it turned out that for a first order phase transition 
the mass of the Higgs particle has to be less then $72$~GeV above which only a
rapid cross-over can be seen. Since the experimental lower bound on the Higgs 
boson mass is much higher, SM baryogenesis is ruled out. In order to avoid
the conclusion that the present baryon number was simply an initial
condition we need to go beyond the Standard Model. The most attractive 
extension
of the SM is the Minimal Supersymmetric Standard Model (MSSM).
In the next chapter I will investigate the possibility of baryogenesis in the
MSSM.

The second signal that I discuss in my thesis comes from the observed
cosmic rays. The highest energy detected cosmic ray events have
macroscopic energy, more than $10^{20}$~eV. It is unlikely that these
particles were accelerated from lower energies to such a high energy.
A more attractive possibility is that they are the decay products of some metastable
superheavy
particle, usually associated with some Grand Unified Theory (GUT).
This appealing possibility clearly points far beyond the SM. In the
third chapter of this thesis I give predictions on the density and energy
scale of the Ultrahigh energy cosmic ray sources.

The MSSM baryogenesis calculations needed lattice simulations of the bosonic
sector of MSSM. This requires a large amount of CPU time. For this purpose a
PC based supercomputer was built at the E\"otv\"os University. In the fourth
chapter I will discuss the hardware and software architecture of this machine.

In the MSSM project I worked together with Ferenc Csikor, Zolt\'an Fodor
P\'al Heged\"us, Tam\'as Herpay, Antal Jakov\'ac and Attila Pir\'oth.
The UHECR calculations were done together with Zolt\'an Fodor while
in the supercomputer project Ferenc Csikor, Zolt\'an Fodor, P\'al Heged\"us,
Viktor Horv\'ath and Attila Pir\'oth were also involved.
This also means that only a part of the results belong to me.
My contribution is the following:
\begin{itemize}
\item{Writing a 5000 line C program for the MSSM lattice simulations; Development
of the heatbath and overrelaxation algorithms for the Higgs and squark
fields.}
\item{Performing finite temperature simulations to determine
the critical point of the Electroweak Phase Transition and then zero
temperature simulations to measure the mass spectrum.}
\item{Determining the phase diagram of MSSM in the $m_U^2$--$T$ plane 
in the infinite volume limit. Finding the bubble wall profile during
the phase transition and measuring the width of the wall and the change of the
ratio of the two Higgs expectation values.}
\item{Determining the density of ultrahigh energy cosmic ray sources based on
the clustering features of observations.}
\item{Finding the $P(r,E,E_c)$ function which gives the probability that
a particle produced with energy $E$ is detected above the energy $E_c$
after propagation over a distance $r$.}
\item{Determining the fragmentation function of the proton at high energies.}
\item{Finding the mass of the superheavy particle that could be 
the source of ultrahigh energy cosmic rays.}
\item{Writing the
job management system and the kernel driver for the communication cards of the
PMS supercomputer.}
\end{itemize}
In my thesis I will concentrate on these aspects of the problems.
I have four publications connected to my thesis:
\begin{itemize}

\item
F.~Csikor, Z.~Fodor, P.~Heged\"us, A.~Jakov\'ac, S.~D.~Katz and A.~Pir\'oth,
Phys.\ Rev.\ Lett.\ {\bf 85}, 932 (2000).

\item
Z.~Fodor and S.~D.~Katz,
Phys.\ Rev.\ D {\bf 63}, 023002 (2001).

\item
Z.~Fodor and S.~D.~Katz,
Phys.\ Rev.\ Lett.\ {\bf 86}, 3224 (2001).

\item
F.~Csikor, Z.~Fodor, P.~Heged\"us, V.~K.~Horv\'ath, S.~D.~Katz and A.~Pir\'oth,
Comput.\ Phys.\ Commun.\ {\bf 134}, 139 (2001).
\end{itemize}

The structure of this thesis is as follows. In chapter \ref{MSSM} I briefly
discuss the possibilities of electroweak baryogenesis in the SM and
MSSM. Then I present the methods used for lattice simulations of the
bosonic sector of MSSM and give the results of finite and zero temperature
simulations. The last two sections of this chapter deal with the phase diagram
of MSSM and the profile of the bubble wall during the phase transition.

Chapter \ref{UHECR} deals with ultrahigh energy cosmic rays. After a short 
introduction the density of sources is determined. The interesting assumption
that the sources of these cosmic rays can be superheavy particles is discussed
in section \ref{sect_gut}.

In chapter \ref{PMS} the PMS supercomputer is described. Both the hardware and
sofware architectures are discussed in detail and results for the performance
are also given.

\chapter{MSSM Baryogenesis} \label{MSSM}
\section{Baryogenesis in the Standard Model}
The world around us is made up of baryonic matter. This fact can be verified by
observations in our vicinity. For distant regions of the Universe we have only
indirect signals. If in some distant segment
of the Universe anti-world domains existed, the annihilation at the world -- anti-world domain walls should
affect the diffuse cosmic gamma-ray background. The observed gamma-ray
spectrum does not seem to support the
possibility of anti-world domains \cite{cohen98}.

The observed asymmetry can either be accepted as an initial condition at the
''Big-Bang'' or explained by some symmetry breaking
mechanism during the evolution of the
Universe. While the first solution is rather simple, the second one gives a
real challenge to particle physics. The final
non-vanishing baryon number of the universe should be derived
from symmetric initial conditions.

Every baryogenesis scenario should fulfill three conditions first stated by
Sakharov \cite{sakharov}:
\begin{enumerate}
\item{Baryon number violation}
\item{C and CP violation}
\item{Departure from thermal equilibrium}
\end{enumerate}
The Standard Model provides an appealing mechanism for baryogenesis. All
three conditions are satisfied at high temperatures (around 100 GeV) where the electroweak
phase transition (EWPT) takes place. Transitions between the different vacua of the
system, called sphalerons,  can change the baryon number, $B$. The sphalerons,
however, do not only change $B$ but also the lepton number $L$ such that
$B-L$ remains constant. This way a $B+L$ asymmetry can be generated. 
The EWPT is the last
possibility during the evolution of the Universe when the baryon number
could be generated \cite{KRS85}. 
In my thesis I will discuss the possibility of electroweak baryogenesis.

Even if during the EWPT all Sakharov conditions are satisfied it does not
still mean that the required baryon number will be generated. If the
sphaleron rate is too high all generated $B+L$ asymmetry will be washed out.
The expansion rate of the universe should be large enough to prevent this.
This condition can be formulated using the expectation value of the Higgs field
($v$) in the symmetry-broken phase and the critical temperature ($T_c$) of
the phase transition:
\bee
\frac{v}{T_c}>1
\ee
The forthcoming part of this chapter will deal with this condition.


The first detailed description of the EWPT in the SM was based on
perturbative techniques \cite{4d_pert}. However, there were ${\cal O}(100\%)$ 
corrections between different orders of the perturbative expansion
for Higgs boson masses larger than about 60 GeV. This questions the use of
perturbation theory in this region. 
The dimensionally reduced 3d effective model (e.g.~\cite{3d_pert}) was also
studied perturbatively and it gave similar conclusions. To solve the problem of
higher Higgs masses lattice simulations were needed. Both direct four
dimensional simulations \cite{4d_latt,4d_latt1} and reduced three dimensional
simulations \cite{3d_latt} were carried out. The results are in agreement
and they contradict perturbation theory. While perturbation theory predicts
a first order phase transition even 
for large Higgs masses, lattice studies show that
the strength of the transition gets weaker and there is an endpoint
\cite{3d_end,4d_end} at Higgs mass $m_H=72\pm1.4$~GeV \cite{4d_end}.

The present experimental lower limit of the SM Higgs boson mass is 
by several standard deviations larger than the endpoint value. Thus any
EWPT in the SM is excluded. This also means that the SM baryogenesis in the
early Universe is ruled out.

\section{Baryogenesis in the MSSM}

In order to explain the observed baryon asymmetry, extended
versions of the SM are necessary. Clearly, the most attractive
possibility is MSSM. According to perturbative predictions the
strength of the EWPT in the MSSM depends strongly
on the scalar masses \cite{mssm_pert}. Since
now there are more scalars than the Higgs itself they can be used to increase
the Higgs mass while keeping $v/T_c$ above $1$.
In particular if the stop mass is smaller than the top mass then baryogenesis
may be possible even for Higgs masses around 100~GeV \cite{light_stop}.
At two-loop level stop-gluon graphs give a considerable
strengthening of the EWPT (e.g.~third and fourth paper of \cite{mssm_pert}).

A reduced 3d version of the MSSM has recently been studied on the
lattice \cite{mssm_3d}. It included
${\mathrm{SU(3)}} \! \times \! {\mathrm{SU(2)}}$
gauge fields, the right-handed stop and a ``light'' combination
of the Higgses. The results show that the EWPT can
be strong enough, i.e.\ $v/T_c \! > \! 1$, up to 
$m_h \! \approx \! 105$ GeV and $m_{\tilde t} \! \approx \! 165$ GeV,
where $m_h$ is the mass of the lightest neutral scalar and
$m_{\tilde t}$ is that of the stop squark.

In this chapter I study the EWPT in the MSSM on four dimensional lattices.
Lattice simulation of fermionic fields is extremely CPU consuming.
Fortunately all problems arising in the perturbative approach come from
the bosonic sector of the theory. Thus we can find a mixed solution:
simulating only the bosonic sector on the lattice, and taking
fermions into account perturbatively \cite{mssm_prl}.
In fact the method we used is to study almost the whole
bosonic sector on the lattice and
then to tune perturbation theory to get the same results. This
''calibrated'' perturbation theory is then used to correct lattice results at
different lattice spacings to be on a line of constant physics (LCP).

Our analysis extends the 3d study \cite{mssm_3d} in two ways:
\begin{enumerate}
\item{ We use 4d lattices instead of 3d. This way the bosonic fields are
directly put on the lattice and the uncertainties
coming from dimensional reduction
are missing.
Using unimproved lattice actions the leading corrections due 
to the finite lattice spacings are proportional to $a$ in 3d and 
only to $a^2$ in 4d. For O($a$) improvement in the 3d case cf. \cite{Moore}. 
In 4d simulations we also have direct control over zero
temperature renormalization effects.}

\item{ We include both Higgs doublets, not only the light combination. 
According to standard baryogenesis scenarios (see e.g. \cite{mssm_gen})
the generated baryon number is connected to the the expectation values and the
relative phase of the two Higgs fields in the bubble wall. I will study the
properties of the bubble wall in section \ref{sect_bubble}.}

\end{enumerate}
In the following sections I will discuss the lattice simulations of the
bosonic sector of MSSM in detail.

\section{MSSM on the lattice}

\subsection{The Lagrangian}

In order to have a reasonably simple Lagrangian while not neglecting important
effects, we decided to ignore scalars with small Yukawa couplins and the U(1)
factor which can easily be treated perturbatively. The fields we kept are the
following:
\begin{itemize}
\item{SU(3) and SU(2) gauge fields of the strong and weak interactions:
$A^{(s)}_\mu$ and $A^{(w)}_\mu$;}
\item{two Higgs doublets: $\ha$ and $\hb$;}
\item{left handed stop-sbottom doublet: $Q_{ij}$ where $i$ is the SU(3)
index while $j$ is the SU(2) index;}
\item{right handed stop and sbottom fields: $U_i$ and $D_i$.}
\end{itemize}

The continuum Lagrangian in standard notation reads
\bee
{\cal L}={\cal L}_g+{\cal L}_k+{\cal L}_V+{\cal L}_{sm}+{\cal L}_Y+
{\cal L}_w+{\cal L}_s.
\ee
The gauge part is
\bee
 {\cal L}_g=\frac{1}{4}\cdot F^{(w)}_{\mu\nu}F^{(w)\mu\nu}+
\frac{1}{4}\cdot F^{(s)}_{\mu\nu}F^{(s)\mu\nu}
\ee
with the usual field-strength tensor.
The kinetic part is the sum of the
covariant derivative terms of all scalars:
\begin{eqnarray}
{\cal L}_k = 
(\DD^{(w)}_\mu \ha)^\dagger (\DD^{(w) \mu} \ha)+
(\DD^{(w)}_\mu \hb)^\dagger (\DD^{(w) \mu} \hb) \nonumber \\
+(\DD^{(ws)}_\mu Q)^\dagger  (\DD^{(ws) \mu} Q)+
(\DD^{(s)}_\mu U^*)^\dagger (\DD^{(s) \mu} U^*)+ 
(\DD^{(s)}_\mu D^*)^\dagger (\DD^{(s) \mu} D^*).
\end{eqnarray}
The potential term for the Higgs fields reads
\begin{eqnarray}
{\cal L}_V= m_{12}^2 [\alpha_1|\ha|^2+ \alpha_2|\hb|^2- 
(\ha^\dagger \tilde{\hb}+ 
h.c.)] \nonumber \\
+ \frac{g_w^2}{8}\cdot (|\ha|^4+ |\hb|^4- 2|\ha|^2|\hb|^2+ 4|\ha^\dagger\hb|^2),
\end{eqnarray}
for which two dimensionless mass parameters are defined:
\bee
\alpha_1=m_1^2/m_{12}^2, \;\;\; \alpha_2=m_2^2/m_{12}^2.
\ee
One gets 
\bee
{\cal L}_{sm}= m_Q^2 |Q|^2+m_U^2 |U|^2+m_D^2 |D|^2
\ee
for the squark mass part, and 
\bee
{\cal L}_Y= h_t^2(|QU|^2+ |\hb|^2|U|^2+ |Q^\dagger\tilde{\hb}|^2)
\ee
for the dominant Yukawa part. 
The quartic parts
containing the squark fields read
\bee
{\cal L}_w= \frac{g_w^2}{8}\cdot [2 \{Q \}^4- |Q|^4+
4|\ha^\dagger Q|^2+ 4|\hb^\dagger Q|^2- 
2|\ha|^2 |Q|^2- 2 |\hb|^2 |Q|^2] 
\ee 
and
\begin{eqnarray}
{\cal L}_s= \frac{g_s^2}{8}\cdot \left[3 \{Q \}^4- |Q|^4+ 2|U|^4+ 2|D|^4-
6|QU|^2 \right. \nonumber \\ 
- 6|QD|^2+ 6|U^\dagger D|^2+ 2|Q|^2|U|^2 
\left. + 2|Q|^2|D|^2- 2 |U|^2|D|^2 \right],
\end{eqnarray}
where 
\bee
\{Q\}^4=Q^*_{i\alpha}Q^*_{j\beta}Q_{i\beta}Q_{j\alpha}.
\ee
The scalar trilinear couplings have been omitted for simplicity.
It is straightforward to obtain the lattice action, for which we used the standard
Wilson plaquette, hopping and site terms.

\subsection{Monte-Carlo techniques}

We used local updates for all fields. The first implementation used the
simplest Metropolis algorithm. However, using this method
${\cal O} (100)$ sweeps were necessary to get independent configurations even
for the smallest lattices. The autocorrelation function which gives the 
correlation between subsequent configurations is defined as:
\bee
a(\tau)=<A_n A_{n+\tau}>-<A_n><A_{n+\tau}>,
\ee
where $A_i$ is the value of some observable on the $i$-th configuration.
The autocorrelation function usually decays exponentially with $\tau$, the
decay rate gives the autocorrelation time. Since this autocorrelation time is 
huge for the Metropolis algorithm, it is
necessary to use faster updating algorithms. For the gauge fields we used the
standard overrelaxation and heatbath updates. For the scalar fields we
had to improve the method used for the simulations of the SU(2)-Higgs model
\cite{4d_latt,4d_latt1}, since the scalar couplings are more complicated.
In the following I describe the overrelaxation and heatbath algorithms for the
scalar fields.

\subsubsection{Overrelaxation algorithm}
The goal of the overrelaxation algorithm is to generate new field
configurations while keeping the action unchanged. In each step only one field at
one lattice site is updated. Let the updated field at a given lattice site be $\phi$.
It is an $N$ component complex vector, where $N=2$ for the Higgs fields and
$N=3$ for the squark fields. The lattice action can be written as:
\bee
S=S_0+\phi^\dagger M \phi +b^\dagger \phi +\lambda |\phi|^4, \label{action_part}
\ee
where $S_0$ is independent of $\phi$, $M$ is an $N \times N$ hermitian matrix,
$b$ is an $N$ component complex vector and $\lambda$ is a scalar. The values of
$M$, $b$ and $\lambda$ are all independent of $\phi$ and can be obtained as
a function of other fields and the given field at other sites.

We would like to update the vector $\phi$ such that $S$ remains unchanged. Due
to the quartic term it is rather difficult to keep $S$ unchanged. However, if
the action changes only slightly then an additional Metropolis accept-reject
step with a high acceptance rate can be performed. If the quartic term vanishes,
the action is invariant under the update:
\bee
\phi \rightarrow 2\cdot\phi_0 -\phi,
\ee
where the action has a local minimum at $\phi=\phi_0$, i.e. $2\cdot M\phi_0 +
b =0$. For $\lambda \neq 0$ this is of course not an exact overrelaxation
step, so it has to be corrected with a Metropolis step. Unfortunately, if the
expectation value of the updated field is large (which is the case for the
Higgs fields in the broken phase), the quartic term becomes large and the
Metropolis acceptance rate gets too low. The algorithm can be improved in the
following way. The action can be rewritten as:
\bee
S=S_0'+\phi^{\dagger} (M+2A\lambda\cdot 1) \phi +b^{\dagger} \phi +
\lambda\left(|\phi|^2-A\right)^2.
\ee
This is clearly equal to (\ref{action_part}), only the constant term ($S_0'$)
has changed. A part of the quartic term was moved to the quadratic term. By
choosing the value of $A$ carefully, the Metropolis acceptance rate can be
significantly increased. The value of $A$ is chosen to be a constant for all
lattice sites and it is tuned to give the best acceptance rate during
thermalization. Using this method we found the
acceptance rate being above 90\% for all scalar fields.

\subsubsection{Heatbath algorithm}

In the heatbath algorithm, the updated $\phi$ is not obtained from its
previous value but is generated directly according to the desired distribution.
The quartic term  makes things again complicated. Just as before,
we can introduce
the $A$ parameter which converts a part of the quartic term to quadratic.
Using an iterative technique the value of $A$ can be set to be equal to 
$|\phi_0|^2$,
where the non-quartic part of the action has its minimum at $\phi_0$:
\bee
2\cdot (M+2A\lambda\cdot 1) \phi_0 +b=0. \label{minimum}
\ee
Let $\chi=\phi-\phi_0$. Then, using \ref{minimum},
the action can be written as:
\bee
S=S_0''+\phi_0^\dagger (M+2A\lambda\cdot 1) \phi_0 +b^\dagger\phi_0+
\chi^\dagger (M+2A\lambda\cdot 1) \chi +
\lambda\left(|\phi_0+\chi|^2-A\right)^2.
\ee
We have to generate $\phi$ according to the distribution $e^{-S}$. This is done
approximately in the following steps. First we find $A$ and $\phi_0$. Then we generate $\chi$ according
to the distribution $\exp(-\chi^\dagger M' \chi)$ with $M'=M+2A\lambda\cdot 1$. This
will not give exactly the required distribution, so we perform an additional
accept-reject step with $dS=\lambda\left(|\phi_0+\chi|^2-A\right)^2$.
The generation of $\chi$ according to the given distribution is fairly simple.
If we decompose $M'$ as $M'=L^\dagger L$ then the required distribution is
$\exp(-|L \chi|^2)$. If $\eta$ is a Gaussian random vector then $\chi=L^{-1} \eta$
will have the desired distribution.

As the generated distribution is not exactly the desired distribution (and hence
is corrected by the accept-reject step), we may call this algorithm as
a ''quasi-heatbath'' algorithm. The acceptance rate for all fields was above 90\%.

\begin{figure}[!htb]
\centerline{\includegraphics*[width=7cm]{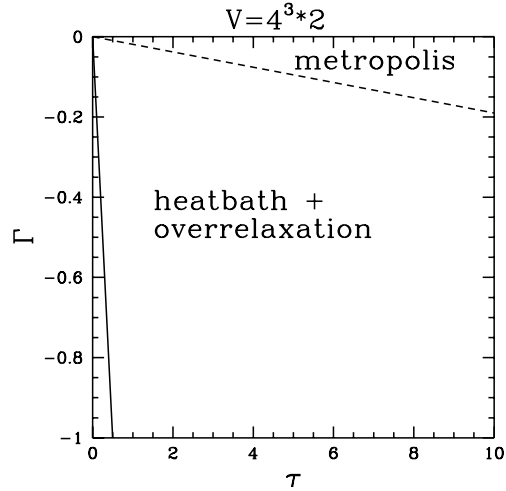}}
\caption[a]{{\sl The autocorrelation function for Metropolis (dashed line)
and
overrelaxation/heatbath (solid line) algorithms. $\Gamma$ is defined as
the logarithm of the autocorrelation function. The exponential decay
can be seen in both cases. The autocorrelation time
is much higher for the Metropolis
algorithm.}}  \label{fig_a_corr}
\end{figure}

The full update is a combination of local overrelaxation and heatbath steps.
Overrelaxation moves fast in the configuration space while heatbath ensures 
ergodicity. We measured the autocorrelation time for this combined updating 
algorithm. Figure \ref{fig_a_corr} shows how much this updating is faster than
the simple Metropolis algorithm for a small $4^3\times2$ lattice. Using 
Metropolis ${\cal O}(100)$ sweeps are required to have a new independent 
configuration, while using the overrelaxation/heatbath combination each 
configuration is practically independent.

\section{Lattice simulations}
The parameter space of the above Lagrangian is many-dimensional. Most
of these parameters must have been fixed.
The experimental values were taken for the strong, weak and Yukawa couplings, 
and $\tan \beta=6$
is used. For the bare soft breaking masses our choice was $m_{Q,D}=250$~GeV, 
$m_U=0$~GeV in the squark sector and $m_{12}=150$~GeV in the Higgs sector. 
The value of $\alpha_1$ was set by the value of $\tan  \beta$ ($\alpha_1
\approx \tan \beta$) while $\alpha_2$ was used to tune the system to the
critical point.

\begin{figure}[!ht]
\centerline{\includegraphics*[width=7cm]{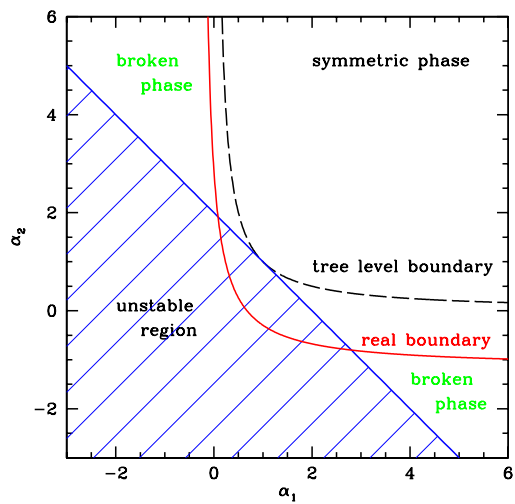}}
\caption[a]{{\sl The schematic phase diagram in the $\alpha_1-\alpha_2$ plane.
The dashed line represents the tree level phase boundary, while the solid line
shows the real boundary. The shaded region is unstable where the action is
not bounded from below.}}  \label{fig_phases}
\end{figure}

Figure \ref{fig_phases} shows the schematic phase diagram of MSSM in the
$\alpha_1-\alpha_2$ plane. We can observe the renormalization effects on
the phase boundary. The tree level boundary is $\alpha_1\cdot\alpha_2=1$
while the real boundary (determined from a few simulation points) has a
similar shape but it is shifted. It is interesting that the phase boundary
is more shifted in the $\alpha_2$ direction than in the $\alpha_1$ direction.
This is caused by the large top Yukawa coupling since it gives a dominant
renormalization contribution to $\alpha_2$ only.

Simulations were performed in two main steps:
\begin{enumerate}
\item{Finite temperature simulations, when the temporal extension $L_t$ of the lattice
is much smaller than the spatial extensions $L_{x,y,z}$. For a given $L_t$
we fixed all parameters of the Lagrangian except $\alpha_2$. We used $\alpha_2$
to tune the system to the transition point. For finding the transition point,
two different methods were used which I will discuss in more details later. We measured
the jump of the expectation value of the Higgs fields in the transition point.
Using lattices with different spatial extensions, we performed
infinite volume extrapolations both for the critical $\alpha_2$ and the Higgs jump.
}
\item{Zero temperature simulations. We used large lattices
(with large $L_t$ as well). $\alpha_2$ was set to its infinite volume extrapolated
value, while all other parameters were set to be 
the same as in the finite
temperature case. We measured the correlation functions for the $W$ and Higgs
bosons and Wilson-loops. The $W$ correlation length gives the inverse
$W$ mass in
lattice units which determines the lattice spacing. This makes
it possible to express our quantities in physical units rather than in lattice
units.}
\end{enumerate}

Both steps were performed for different lattice spacings. The lattice
spacing can be changed by changing the temporal extension of the finite
temperature lattice. We used four different temporal extensions,
$L_t=2,3,4,5$.
While changing the lattice spacing, we have to change the bare parameters
of the Lagrangian so that all zero temperature observables
remain the same. We want to keep the system on a line of constant physics (LCP).

\subsection{Finite temperature simulations} \label{sect_finT}

The first goal of finite temperature simulations is to determine the
critical value of $\alpha_2$ at which a first order phase transition
occurs. Depending on the strength of the phase transition, different methods
can be efficiently used to find the value of $\alpha_{2c}$ accurately.
First one can get a rough estimate on $\alpha_{2c}$ if simulations are
performed for several different $\alpha_2$ values. A change can be observed
in the length of the second Higgs field $|H_2|^2$ around the critical point.
Then collecting large statistics near the critical point, the
Ferrenberg-Swendsen reweighting technique \cite{FerSwen} can be applied
to get information about observables belonging to slightly different
$\alpha_2$ values. We used two different methods to find the critical point.

\subsubsection{Lee-Yang zeros}

The method of Lee-Yang zeros \cite{LeeYang} can be applied if the phase
transition is not too strong and during the simulations the system can
walk between the different phases.

The free energy is singular in a first order phase transition point, which
means that the partition function vanishes. For finite volumes the free
energy singularity, the roots of the partition function will not be real any
more. If we analytically continue the partition function into
the complex $\alpha_2$ plane, we can find these zeroes.
Fortunately as we increase the lattice volume, the imaginary parts of these
Lee-Yang zeros tend to be zero. The partition function at arbitrary $\alpha_2$
(not too far from the simulation point) can be obtained by the 
reweighting mentioned before. We may then look for the zeros of the partition
function in the complex plane and the real part of the root with smallest
imaginary part will give the critical point to a high accuracy.

\subsubsection{Constrained simulations}

Lee-Yang zeros can not be found (or fake zeros are found)
if the phase transition is too strong. In these
cases the system is usually stuck into one of the phases and due to
the ''supercritical slowing down'' the updating algorithm is not able to evolve it
to the other phase.

A solution to this problem is not to let the system go to any of the
phases. Instead, we constrain the value of the order parameter (in our
case $|H_2|^2$ averaged over the lattice) in a small
interval between the two phases \cite{Bhanot}. All configurations that have an order parameter
out of this interval are rejected. If we are at the critical point
then the system can minimize its free energy in a way that two separate
phases are formed at two different parts of the lattice.
The definition of the critical point is  in this case that the distribution
of the order parameter on the selected interval should be uniform. This 
definition may slightly depend on the choice of the interval between the
two phases, however, this dependence vanishes in the infinite volume limit.

\subsubsection{Finite temperature results}

\begin{figure}[!ht]
\centerline{\includegraphics*[width=7cm]{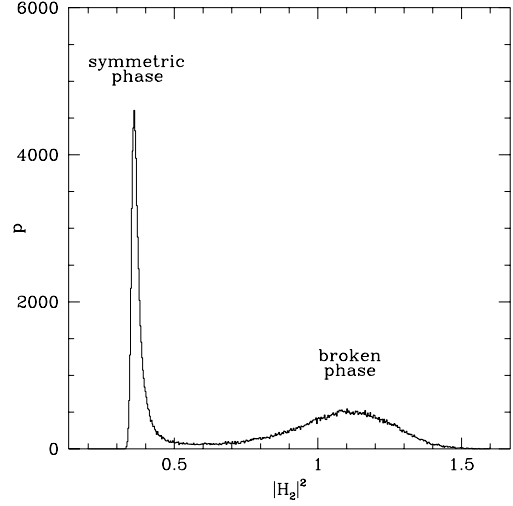}}
\caption[a]{{\sl The distribution of the order parameter $|H_2|^2$ in
the transition point. The two separate peaks clearly indicate a first order
phase transition.}}\label{fig_two_peaks}
\end{figure}

In the following I present the results of finite temperature simulations.
Four different temporal lattice extensions were used ($L_t=2,3,4,5$).
Figure \ref{fig_two_peaks} shows the distribution of the order parameter
in the transition point on a $4^3\times 2$ lattice. It is easy
to identify the two phases, the transition is clearly of first order.

\begin{figure}[!ht]
\centerline{\includegraphics*[width=7cm]{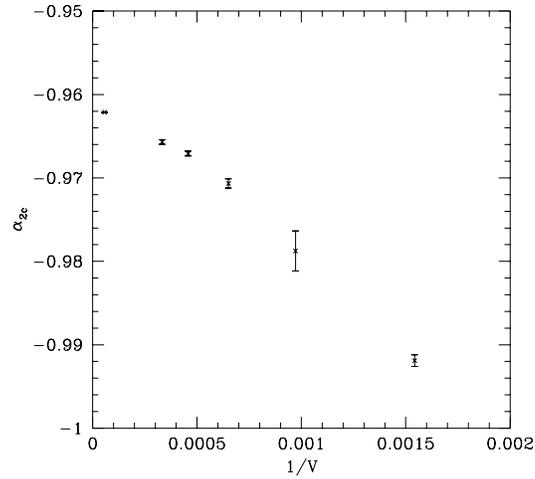}}
\caption[a]{{\sl The critical $\alpha_2$ values for different lattice volumes
as a function of the inverse volume.}} \label{fig_scaling}
\end{figure}

Using the method of Lee-Yang zeros, the critical $\alpha_2$ value was found
for several spatial lattice extensions. Figure \ref{fig_scaling} shows the
finite volume scaling of critical points for $L_t=3$. One can observe that
the critical points scale well with $1/V$ (except for the smallest volumes)
and this property can be used to
extrapolate to infinite volume. The infinite volume $\alpha_2$ is
found to be $\alpha_{2c}=-0.96137(3)$. The jump of the order parameter can
be measured from histograms like Figure \ref{fig_two_peaks} and an infinite
volume extrapolation can also be done.

Table \ref{tab_T} shows $\alpha_{2c}$ and $v/T$ for the four different temporal
extensions. The errors were obtained by using a jackknife analysis. We can
observe that both $\alpha_{2c}$  and the Higgs jump increases
as we go to the continuum limit.

\begin{table}[!ht]
\begin{center}\begin{tabular}{c|c|c|c|c}
$L_t$ 		& $2$ 	& $3$ 	& $4$ 	& $5$ 	\\
\hline
$\alpha_{2c}$ 	&-1.0005(5)	&-0.96137(3)	&-0.9575(1) 	&-0.95504(2) \\
\hline
$v/T$ 		&1.50(1) &1.83(5)	&1.99(2)	&2.12(1) 	\\
\end{tabular}
\caption[a]{\label{tab_T}
{ \sl The critical $\alpha_2$ values and $v/T$ for different
temporal lattice extensions.
}}
\end{center}\end{table}

\subsection{Zero temperature simulations}
After performing finite temperature simulations and finding the critical
$\alpha_2$ values we have to carry out zero temperature simulations to
measure the mass spectrum at the given physical point. The masses of
particles can be extracted from two-point functions of field
operators. For the Higgs mass we measured the correlation functions:
\bee
c_{ij}(x,y)=<|H_i(x)|^2|H_j(y)|^2>,
\ee 
with $i,j=1,2$ for the two Higgs fields.
The lightest Higgs mass can be obtained from the exponential tails of
these correlation functions. For the $W$ mass the same operators were
used as in \cite{4d_latt1}. The results for the Higgs and $W$ masses and their
ratio $R_{HW}$ is given in Table \ref{tab_T0}. The errors were
calculated again by a jackknife analysis. We can see from the data that not an exact
line of constant physics was followed. Thus we need perturbative corrections
which we include in the next section.

\begin{table}[!ht]
\begin{center}\begin{tabular}{c|c|c|c|c}
$L_t$ 		& $2$ 	& $3$ 	& $4$ 	& $5$ 	\\
\hline
$m_H$ 	 	&0.325(6)&0.124(10) 	&0.088(6) &0.070(10) 	\\
\hline
$m_W$ 		&0.594(12)&0.335(8)	&0.253(18)&0.211(20) 	\\
\hline
$R_{HW}$       	&0.547(15)&0.370(31)	&0.348(34)&0.332(57) 	\\
\end{tabular}
\caption[a]{\label{tab_T0}
{ \sl The lightest Higgs and $W$ masses in lattice units and their ratio
for different temporal lattice extensions. $L_t$ is the temporal lattice
extension of the corresponding finite temperature lattice.
}}
\end{center}\end{table}

\section{Comparison with perturbation theory}

We compared our simulation results with perturbation theory.
We used one-loop perturbation theory without applying high
temperature expansion (HTE). A specific feature was a careful treatment 
of finite renormalization effects, by 
taking into account all renormalization corrections and adjusting them 
to match the measured zero temperature spectrum  
\cite{mssm_prd}. We studied also the effect of the
dominant finite temperature two-loop diagram (``setting-sun'' stop-gluon
graphs, cf.\ fifth ref.\ of \cite{mssm_pert}), but only in the HTE. 
Since the infrared behavior of the setting-sun graphs is not fully
understood, we used the one-loop technique with the zero temperature
scheme defined 
above.
This type of one-loop perturbation theory was also applied to correct
the measured data to some fixed LCP quantities, which are
defined as the averages of results at different lattice spacings, (i.e.\ our 
reference point, for which the most important quantity is the lightest 
Higgs mass).
  
The bare squark mass parameters $m_Q^2,m_U^2,m_D^2$ receive quadratic 
renormalization corrections. As it is well-known, one-loop lattice perturbation
theory is not sufficient to determine these corrections reliably, 
thus we used the following method. 
We first determined the position of the non-perturbative color-breaking
phase transitions in the bare quantities (see next section).
These quantities were compared with the prediction of the
continuum perturbation theory,
which yielded the renormalized mass parameters on the lattice.  

\begin{figure}[!ht]
\centerline{\includegraphics*[width=7cm]{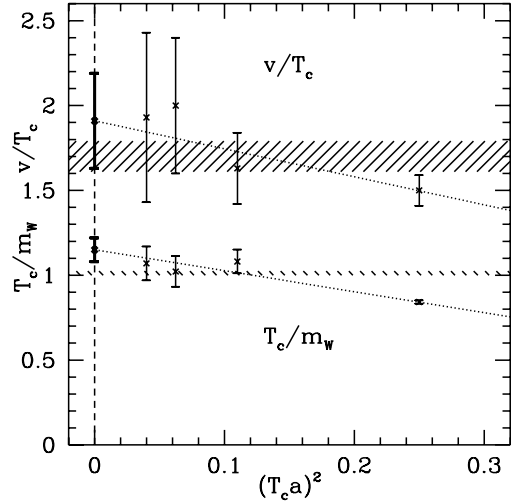}}
\caption[a]{{\sl The normalized jump and the critical temperature
in the continuum limit.}}\label{jump}
\end{figure}

Figure~\ref{jump} contains the continuum limit extrapolation for
the normalized jump of the order parameter ($v/T_c$: upper data)
and the critical temperature ($T_c/m_W $: lower data). The shaded
regions are the perturbative predictions at our reference point 
(see above) in the continuum. Their widths reflect 
the uncertainty of our reference point, which is dominated by the error of 
$m_h$. Note that $v/T_c$ is very sensitive to $m_h$, which results in the 
large uncertainties. Results obtained on the lattice  and in
perturbation theory agree reasonably within the estimated uncertainties.
It might well be that the $L_t$=2 results are not in the scaling region;
leaving them out from the continuum extrapolation the agreement between 
the lattice and perturbative results is even better.

\section{The phase diagram of bosonic MSSM}

As the bare squark mass parameter $m_U^2$ is decreased, the system can go
to a so-called color-breaking phase where the SU(3) symmetry is spontaneously broken.
As mentioned in the previous section, the knowledge of this
well-defined transition point helps us to perform the squark mass
renormalization. We determined the phase diagram in the $m_U^2$--$\alpha_2$
plane in the infinite volume limit 
and after performing zero temperature simulations
we could transform it to the physical $m_U^2$--$T$ scales. The phase
transition to the color-breaking phase is much stronger than the one between
the symmetric and Higgs phases, so we used the constrained simulation method
to obtain the critical $m_U^2$ values.

\begin{figure}[!ht]
\centerline{\includegraphics*[width=7cm]{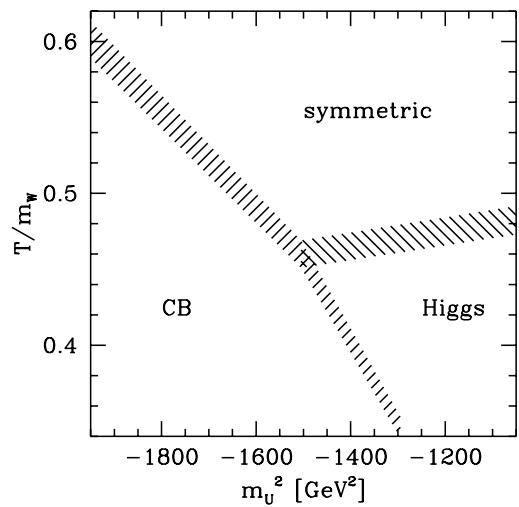}}
\caption[a]{{\sl The phase diagram of the bosonic theory obtained by lattice 
simulations.}}\label{phase}
\end{figure}

Figure~\ref{phase}. shows the phase diagram in the
$m_U^2$--T plane. One can identify three phases. The phase on the left
(large negative $m_U^2$ and small stop mass) is the
color-breaking  phase. The phase in the upper right part is the
symmetric phase, whereas the Higgs phase can be found in the
lower right part.  The line separating the symmetric and Higgs phases
was obtained from the $L_t=3$ simulations, whereas the lines between these
phases and the  color-breaking 
one were determined by keeping the lattice
spacing fixed, while increasing and decreasing the temperature by
changing $L_t$ to 2 and 4, respectively. The shaded regions indicate
the uncertainty in the critical temperatures. The qualitative features of
this picture are in complete agreement with perturbative and 3d
lattice results \cite{mssm_pert,light_stop,mssm_3d}; however, our choice of
parameters does not correspond to a two-stage symmetric-Higgs phase 
transition. In this 
two-stage scenario there is a phase transition from the symmetric to
the color-breaking phase at some $T_1$ and another phase
transition occurs at $T_2<T_1$ from the color-breaking to the Higgs phase.  
It has been argued \cite{cline99} that in the early universe no
two-stage phase transition took place.

\section{Analysis of the bubble wall} \label{sect_bubble}

In order to produce the observed baryon asymmetry, a strong first order
phase transition is not enough.
According to standard MSSM baryogenesis scenarios \cite{mssm_gen} the generated
baryon asymmetry is directly proportional to the variation
of $\beta$ through the bubble wall separating the Higgs and symmetric
phases. 
By using elongated lattices ($2\cdot L^2 \cdot 192$), $L$=8,12,16 
at the transition point we studied
the properties of the wall. We performed constrained simulations
in the transition point (see section \ref{sect_finT}). The length of the
second Higgs field was restricted
to a small interval between its values in the bulk 
phases. As a consequence, the system fluctuated
around a configuration with two bulk phases and 
two walls between them. In order to have the smallest
possible free energy, the walls are perpendicular to the 
long direction. To find the wall profile we measured the average of
both Higgs fields on each timeslice separately. When taking the averages
of the wall profiles one has to be careful. Due to translational symmetry
the location of the walls is different for different configurations. Before
taking an average the different configurations should be appropriately shifted.
The required shift between two samples
can be obtained by minimizing the following ''distance'':
\bee
d=\sum_{i=1}^{192}{\frac{\left(\Phi_{1i}-\Phi_{2(i+k)}\right)^2}{\sigma_{1i}^2+\sigma_{2(i+k)}^2}},
\ee
where $\Phi_{1i}$ ($\Phi_{2(i+k)}$) is the average of one of the Higgs fields
on the
$i$-th ($i+k$-th) timeslice on the first (second) sample and the $\sigma$-s
are the corresponding standard deviations. $k$ is the relative shift of the two samples
($i+k$ is of course meant by modulo 192).

\begin{figure}[!ht]
\centerline{\includegraphics*[width=6.8cm]{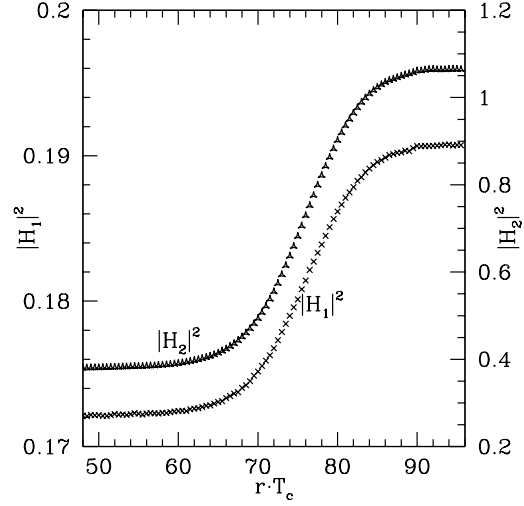}}
\caption[a]{{\sl The profile of the bubble wall for both of the Higgs fields 
for the lattice $2\cdot 12^2 \cdot 192$.}}
\label{fig_wall}
\end{figure}

Figure~\ref{fig_wall} shows the bubble wall profiles for both Higgs fields after
averaging ${\cal O}(50000)$ configurations for $L=12$. The wall profiles
are similar for $L=8$ and $L=16$. The width of the wall ($w$) can be obtained
by fitting $a+b\cdot \tanh(\frac{2(x-x_0)}{w})$ to the wall profile. 
The width slightly depends on the cross size of the lattice. We
found
$w=[A+B\cdot \log (aLT_c)]/T_c$ with $A=10.8\pm.1$ and $B=2.1\pm.1$. 
This behavior indicates that the bubble wall is
rough and without a pining force of finite size its width diverges
very slowly (logarithmically) \cite{Jasnow}.
For the same bosonic theory the perturbative approach predicts 
$(11.2\pm1.5)/T_c$ for the width. 

\begin{figure}[!ht]
\centerline{\includegraphics*[width=6.8cm]{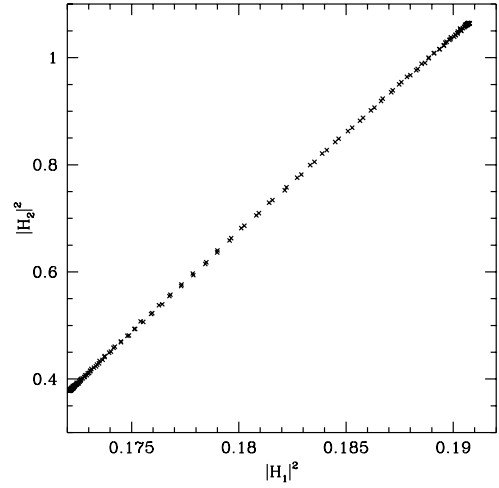}}
\caption[a]{{\sl The relation between the Higgs fields within the bubble wall. The
variation of $beta$ through the wall can be determined by de slopes at the
two ends.}}
\label{fig_tanbeta}
\end{figure}

Transforming the data
of Figure~\ref{fig_wall} to $|H_2|^2$ as a function of $|H_1|^2$, 
we obtain Figure \ref{fig_tanbeta}. We can see that the relation between
the lengths of the Higgs fields is almost linear. However, the slopes at
the two ends significantly differ. From this difference we can compute the
variation of $\beta$ through the bubble wall:
$\Delta \beta=0.0061\pm 0.0003$.
The perturbative prediction at this point is $0.0046\pm0.0010$.
Thus perturbative studies such as \cite{wall_pert} 
are confirmed by non-perturbative results.

The errors of the wall width and $\Delta \beta$ were found by a jackknife
analysis.

\chapter{Ultrahigh energy cosmic rays} \label{UHECR}

The existence of Ultrahigh energy cosmic rays (UHECRs) -- those with energy above
$10^{20}$~eV -- is a real challenge for conventional theories of their origin
based on acceleration of charged particles.

The current theories of UHECR can be divided into two broad classes: the
''bottom-up'' and ''top-down'' scenarios. They are opposite to each other.
In the ''bottom-up'' scenarios it is assumed that UHECRs are accelerated
from lower enegries in special astrophysical environments.
Some examples are acceleration in shocks associated with supernova
remnants, 
active galactic nuclei (AGNs), powerful radio galaxies, 
or acceleration in the strong electric fields generated
by rotating neutron stars. 
In the ''top-down'' scenarios UHECRs are the decay products of some yet 
unknown metastable superheavy particles. 

If these UHECRs are conventional particles such as nuclei or protons,
then above energies of $4\times 10^{19}$~eV they loose a large fraction
of their energy due to the Greisen-Zatsepin-Kuzmin (GZK) effect \cite{GZK66}.
This predicts a sharp drop in the cosmic ray flux above
the GZK cutoff around $4\cdot 10^{19}$~eV. The 
available data shows no such drop. About 20 events above $10^{20}$~eV
were observed by a number of experiments such as AGASA 
\cite{AGASA}, Fly's Eye \cite{FLY}, Haverah Park \cite{HAVERAH},
Yakutsk \cite{YAKUTSK} and HiRes \cite{HIRES}. Since above
the GZK energy the attenuation length of particles is a few tens
of megaparsecs \cite{YT93,BS00,AGNM99,SEMPR00},
if an UHECR is
observed on earth it must be produced in our vicinity (except for UHECR
scenarios based on weakly interacting particles, e.g. neutrinos  \cite{DKDM00}).

Usually it is assumed that at these high energies the galactic
and extragalactic magnetic 
fields do not affect the orbit of the cosmic rays, thus they 
should point back to their origin within a few degrees. In 
contrast to the low energy cosmic rays one can use UHECRs
for point-source search astronomy. (For an extragalactic
magnetic field of $\mu$G rather than the usually assumed nG 
there is no directional correlation with the source \cite{FP00}.) 

Though there are some peculiar clustered events \cite{Hea96,Uea00}, which we discuss in 
detail in the next section \cite{FK00},
the overall distribution of UHECR on the sky is practically isotropic \cite{BM99}. 
This observation is rather surprising since in principle only a few 
astrophysical sites (e.g. active galactic nuclei \cite{M95} or the extended 
lobes of radio galaxies \cite{RB93}) are capable of  accelerating such 
particles. Nevertheless none of the UHECR events came from these 
directions \cite{ES95}.
Sources of extragalactic origin (e.g. AGN \cite{BS87}, topological
defects \cite{HSW87} or the local supercluster \cite{BG79}) should 
result in a GZK cutoff, which is in disagreement with experiments.  
Hence it is generally believed \cite{B99} that there is no 
conventional astrophysical explanation for the observed UHECR spectrum.

The ''top-down'' scenarios are other candidates to explain the
highest energy events. This possibility will be discussed
in section \ref{sect_gut}. I will concentrate on
the determination of the mass scale of the superheavy $X$ particle.
The existence of these $X$ particles would clearly point beyond the Standard
Model and the most interesting result is that the $m_X$ obtained using the
experimental data is consistent with the GUT scale \cite{FK01}.

\section{Clustering of UHECR events}
The arrival directions of 
the UHECRs measured by experiments show some peculiar clustering: 
some events are grouped within $ \sim 3^o$, the typical angular 
resolution of an experiment. Above $4\cdot 10^{19}$ eV 92 cosmic ray events 
were detected, including 7 doublets and 2 triplets. 
Above $10^{20}$ eV one doublet out of 14 events was found \cite{Uea00}. 
The chance probability of such a clustering from uniform distribution is 
rather small \cite{Uea00,Hea96}. (Taking the average bin $3^o$
 the probability of generating one doublet out of 14 events is $11\%$.)

The clustered features of the events initiated
an interesting statistical analysis  
assuming compact UHECR sources \cite{DTT00}. The authors found
a large number, $\sim 400$ for the number of 
sources\footnote{Approximately $400$ sources
within the GZK sphere results
in one doublet for 14 events. The order of magnitude of this result is
in some sense 
similar to that of a ``high-school'' exercise: what is the minimal 
size of a class for which the probability of having clustered birthdays
--at least two pupils with the same birthdays-- is larger than 50\%. 
In this case the number of ``sources'' is the number of possible 
birthdays $\sim 400$. In order to get the answer 
one should solve $365!/[365^k (365-k)!] < 0.5$, which gives as a minimal size
k=23.}
inside a GZK sphere of 25~Mpc. 
They assumed that \hfill\break
a.) the number of clustered events is much smaller than the total 
number of events (this is a reliable assumption at present
statistics; however, for any number of sources the increase of
statistics, which will happen in the near future, results in more 
clustered events than unclustered);\hfill\break
b.) all sources have the same luminosity which gives a delta function 
for their distribution (this unphysical choice represents an important 
limit, it gives the smallest source density for a given
number of clustered and unclustered events).\hfill\break
c.) The GZK effect makes distant sources fainter; however, this feature
depends on the injected energy spectrum 
and the attenuation lengths and elasticities of the propagating
particles. In \cite{DTT00} an exponential decay was used with an energy
independent decay length of 25Mpc.

In our approach none of these assumptions were used. In addition
we included spherical astronomy corrections and in particular
determined
the upper and lower bounds for the source density
at a given confidence level. As it will be shown, the most probable value
for the source density is really large; however, the 
statistical significance of this result is rather small. At
present the small number of UHECR events allows a 95\%
confidence interval for the source density which spreads 
over four orders of magnitude. Since future experiments,
particularly Pierre Auger \cite{PAUG}, will have a much
higher statistical significance on clustering (the expected 
number of events of $10^{20}$ eV and above is 60 per year 
\cite{BBL00}), we present our results on the density of sources also 
for larger number of UHECRs above $10^{20}$ eV.

In order to avoid the assumptions of \cite{DTT00} a combined analytical 
and Monte-Carlo technique will be presented adopting the 
conventional picture of protons as the ultrahigh energy cosmic rays.
Our analytical approach of Section \ref{sect_anal}
gives the event clustering 
probabilities for any space, luminosity and energy distribution of 
the sources by using a single additional function $P(r,E;E_c)$, the probability
that a proton created at a distance $r$ with energy $E$ arrives
at earth above the threshold energy $E_c$ \cite{BW99}. 
With our Monte-Carlo technique of Section \ref{sect_monte} we
determine the probability function $P(r,E,E_c)$ for 
a wide range of parameters. 

\subsection{Analytical approach} \label{sect_anal}

The key quantity for finding the distribution functions for the
source density, is the probability of detecting $k$ events from one randomly
placed source. The number of UHECRs emitted by a source of $\lambda$ luminosity
during a period $T$ follows the Poisson distribution. 
However, not all
emitted UHECRs will be detected. They might loose their energy during
propagation or can simply go to the wrong direction.

For UHECRs the energy loss
is dominated by the pion production in interaction with
the cosmic microwave background radiation. In ref.
\cite{BW99} the probability function $P(r,E,E_c)$ was presented
for three specific threshold energies. This function gives 
the probability that a proton
created at a given distance from earth (r) with some energy (E) is detected
at earth above some energy threshold ($E_c$). The resulting 
probability distribution can be approximated over the energy range 
of interest by a function of the form
\begin{equation}\label{bahcall}
P(r,E,E_c)\approx \exp[-a(E_c)r^2\exp(b(E_c)/E)]
\end{equation}
The appropriate values of $a$ and $b$ for $E_c/(10^{20}{\rm eV})=$1,3, and 
6 are, respectively $a/(10^{-4}{\rm Mpc}^{-2})=$1.4, 9.2 and 11,
$b/(10^{20}eV)=$2.4, 12 and 28.

For the sources we use the
second equatorial coordinate system: ${\bf
x}$ is the position vector of the source characterized by ($r,\delta,\alpha$)
with $\delta$ and $\alpha$ being the declination and right ascension,
respectively. The features of the Poisson distribution enforce us to take
into account the fact that the sky is not
isotropically observed. There is a circumpolar cone, in which the sources
can always be seen, with half 
opening angle $\delta'$ ($\delta'$ is the declination of the detector,
for the experiments we study $\delta' \approx 40^o-50^o$). There is also an
invisible region with the same opening angle. Between them there is a region
for which the time fraction of visibility, $\gamma(\delta,\delta')$ is a 
function of the declination of the source. It is straightforward to
determine $\gamma(\delta,\delta')$ for any $\delta$ and $\delta'$:
\begin{equation}
\gamma(\delta,\delta')=\left\{ 
\begin{array}{lll}
0\  &\mbox{if }& -\pi/2< \delta \leq \delta'-\pi/2 \\
1-&\arccos&(\tan\delta'\tan\delta)/\pi \\
 &\mbox{if }& \delta'-\pi/2 <\delta
\leq \pi/2-\delta' \\
1\ &\mbox{if }& \pi/2-\delta' < \delta \leq \pi/2
\end{array}
\right.
\end{equation}
To determine the probability that a particle arriving from random direction
at a random time is detected
we have to multiply $\gamma(\delta,\delta')$ by the cosine of the
zenith angle $\theta$.
In the
following we will use the time average of this function:
\begin{equation}
\eta(\delta,\delta')=\frac{1}{T}\int _0^T{\gamma(\delta,\delta')\cdot
\cos\theta(\delta,\delta',t) dt}
\end{equation}
Since $\delta'$ is constant, in the rest of the paper we do not indicate 
the dependence on it. Neglecting these spherical astronomy 
effects means more than
a factor of two for the prediction of the source density.

The probability of detecting $k$ events from a source at distance 
$r$ with energy $E$ can be obtained by including
$P(r,E,E_c) A\eta(\delta)/(4\pi r^2)$ in the Poisson distribution:
\begin{eqnarray}
p_k({\bf x},E,j)
=\frac{\exp\left[  -P(r,E,E_c)\eta(\delta)j/r^2 \right] }{k!}\times \nonumber\\
\left[ P(r,E,E_c)\eta(\delta)j/r^2\right] ^k, \label{poiss2}
\end{eqnarray}
where we introduced $j=\lambda T A/(4\pi)$ and $A\eta(\delta)/(4\pi r^2)$ 
is the 
probability that an emitted UHECR points to a detector of 
area $A$. 
We denote the space, energy and 
luminosity  distributions of the sources by $\rho({\bf x})$, 
$c(E)$ and $h(j)$, respectively. The probability of detecting $k$
events above the threshold $E_c$ from a single source 
randomly positioned within a sphere of radius $R$ is
\begin{eqnarray}\label{P_k}
P_k=\int_{S_R} dV\; \rho({\bf x}) \int_{E_c}^{\infty} 
dE\; c(E) \int_0^{\infty} dj\; h(j) \times \nonumber \\ 
\frac{\exp\left[ -P(r,E,E_c)\eta(\delta)j/r^2\right] }{k!} \left[
P(r,E,E_c)\eta(\delta)j/r^2 \right] ^k. \end{eqnarray}

Denote the total number of sources within the sphere of 
sufficiently large radius (e.g. several times the GZK radius)
by $N$ and the number of sources that gave $k$ detected events by
$N_k$. Clearly, $N=\sum_0^{\infty}N_i$ and the total number of detected
events is $N_e=\sum_0^{\infty}i N_i$. The probability that for $N$
sources the numbers of different detected multiplets are $N_k$ is:
\begin{equation}\label{distribution}
P(N,\{N_k\})=N!\prod_{k=0}^{\infty} \frac{1}{N_k!}P_k^{N_k}.
\end{equation}
The value of $P(N,\{N_k\})$ is the most important quantity 
in our analysis of UHECR clustering. For a given set of 
unclustered and clustered events ($N_1$ and 
$N_2,N_3$,...) 
inverting the $P(N,\{N_k\})$ distribution function 
gives the most probable value for the number 
of sources and also the confidence interval for it.
If we want to determine the density of sources we can take the limit
$R \rightarrow \infty, N \rightarrow \infty$, while the density of
sources $S=N/(\frac{4}{3}R^3\pi)$ is constant.

\begin{figure}[!ht]
\begin{center}
\includegraphics*[width=8.0cm]{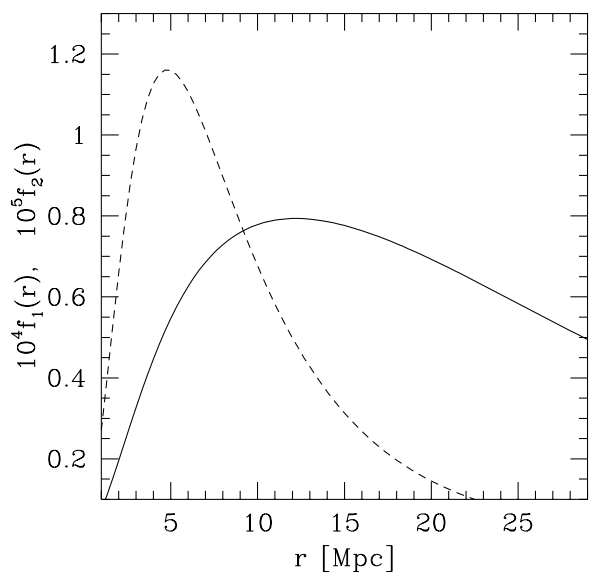}
\caption[a]{\label{f_r}
{ \sl The distributions $f_1(r)$ --solid line-- and $f_2(r)$
--dashed line-- of eqn. (\ref{dominance}). The singlet 
and doublet events are dominated by distance scale of 10-15 Mpc
and 3-5 Mpc, respectively.
}}
\end{center}\end{figure}

In order to illustrate the dominant length scale it is instructive 
to study the integrand $f_k(r)$ of the distance integration in 
eqn. (\ref{P_k}) 
\[
P_k=\int_0^R \left(\frac{dr}{R}\right) f_k(r),
\]
\begin{eqnarray}\label{dominance}
f_k(r)=R r^2 \int d\Omega \rho({\bf x}) \int_{E_c}^{\infty} 
dE\; c(E) \int_0^{\infty} dj\; h(j) \times \nonumber \\ 
\frac{\exp \left[ -P(r,E,E_c)\eta(\delta)j/r^2\right] }{k!}
\left[ P(r,E,E_c)\eta(\delta)j/r^2 \right] ^k.
\end{eqnarray}
Figure \ref{f_r} shows that $f_1(r)$, which leads to singlet events, 
is dominated by the distance scale of 10-15 Mpc, whereas
$f_2(r)$, which gives doublet events, is dominated by the distance
scale of 4-6 Mpc. 
It is interesting that the dominant distance scale for singlet events is
by an order of magnitude smaller than the attenuation length of the protons
at these energies ($l_a\approx 110$~Mpc). This surprising result can be
illustrated using a simple approximation. Assuming that the probability of
detecting a
particle coming from distance $r$ is proportional to $exp(-r/l_a)/r^2$,
$P_1$ will be proportional to $\int d\Omega dr r^2\cdot
\exp[-j\exp(-r/l_a)/r^2]\cdot \exp(-r/l_a)/r^2$. For the typical $j$ values
the $r$ integrand has a maximum around 4 Mpc and not at $l_a$.
These typical distances partly justify our assumption
of neglecting magnetic fields.
The deflection of singlet events
due to magnetic fields does not change the number of multiplets,
thus our conclusions remain unchanged. The typical distance for higher 
multiplets is quite small, therefore deflection can be
practically neglected. Clearly, the fact that multiplets
are coming from our ``close'' neighborhood does not mean
that the experiments reflect just the densities of these
distances. The overwhelming number of events are singlets
and they come from much larger distances. Note, that these 
$f_1(r)$ and $f_2(r)$ functions were obtained with our
optimal $j_*$ value (cf. Figure \ref{ell} and explanation there and in
the corresponding text).
Using the largest possible $j_*$ value allowed by
the 95\% confidence region the dominant distance scales
for $f_1(r)$ and $f_2(r)$ functions turn out to be
30 Mpc and 20 Mpc, respectively.

Note, that $P_k$ and then $P(N,\{N_k\})$ are easily determined by
a well-behaved four-dimensional numerical integration
(the $\alpha$ integral can be factorized)
for any $c(E)$, $h(j)$ and $\rho (r)$ distribution functions
. 
In order to illustrate the uncertainties and sensitivities of the
results we used a few different 
choices for these distribution functions. 

For $c(E)$ we studied three possibilities. The most 
straightforward choice is the extrapolation of the `conventional
high energy component' $\propto E^{-2}$. Another possibility is
to use a stronger
fall-off of the spectrum at energies just below the GZK cutoff,
e.g. $\propto E^{-3}$. These choices span the range usually 
considered in the literature and we will study both of them.
The third possibility is to assume that topological defects generate
UHECRs through production of superheavy particles \footnote{Note, that 
these particles are not superheavy dark matter
particles \cite{ELN90}, which are located
most likely in the halo of our galaxy. These superheavy dark matter
particles can
also be 
considered as possible sources of UHECR \cite{BKV97,BS98,BEA} with anisotropies 
in the arrival direction \cite{PBRS}.}.
According to \cite{BS98}
these superheavy particles decay into quarks and gluons which initiate 
multi-hadron cascades through gluon bremsstrahlung. These finally hadronize
to yield jets.
The energy spectrum was first calculated in \cite{Hill} for the
Standard Model and in \cite{BK98} for the Minimal Supersymmetric 
Standard Model. In Section \ref{sect_gut} we will determine the decay spectrum
of $X$ particles and find the $m_X$ for which the agreement with
observations is the best.
We used this spectrum as the third choice of energy distribution, $c(E)$.

In ref. \cite{DTT00} the authors have shown that for a fixed
set of multiplets the minimal density of sources can be obtained 
by assuming a delta-function distribution for $h(j)$. We 
studied both this limiting case ($h(j)=\delta(j-j_*)$) and a more realistic one
with Schechter's luminosity function \cite{Schechter}:
\begin{equation} \label{spread}
h(j)dj=h\cdot (j/j_*)^{-1.25}\exp(-j/j_*)d(j/j_*). 
\end{equation} 

The space distribution of sources can be given based on some
particular survey of the distribution of nearby galaxies 
\cite{WFP97} or on a correlation length $r_0$ characterizing 
the clustering features of sources \cite{BW99}. For simplicity 
the present analysis deals with a homogeneous distribution of
sources randomly scattered in the universe (Note, that due to the Local 
Supercluster the isotropic distribution is just an approximation.).

\begin{figure}[p]\begin{center}
\includegraphics[width=7.7cm]{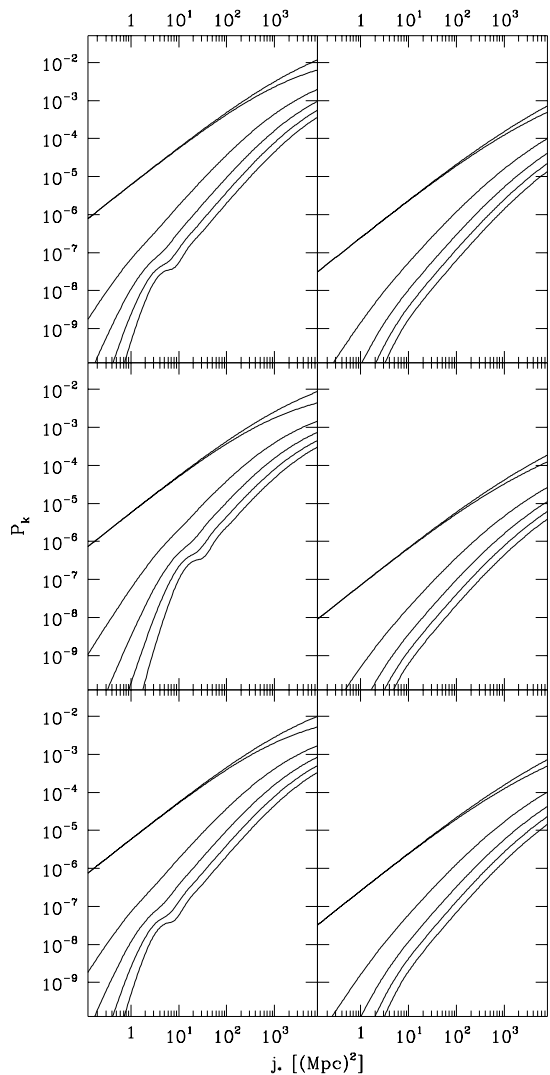}
\caption[a]{\label{p_k}
{ \sl 
The individual $P_k(j_*)$ functions for the different $c(E)$ and 
$h(j)$ choices. The column on the left corresponds to the 
Dirac-delta distribution $h(j)=\delta(j-j_*)$, 
whereas the column on the right 
shows the results for Schechter's luminosity distribution.
The first, second and 
third rows correspond to the $c(E)$ functions proportional to
$E^{-2}$, $E^{-3}$ and the superheavy decay mode, respectively
(see text). On each panel the individual lines from top to 
bottom are: $1-P_0$, $P_1$, $P_2$, $P_3$, $P_4$ and $P_5$.  
}}
\end{center}\end{figure}

Figure \ref{p_k} shows the resulting $P_k(j_*)$ probability functions for the 
different choices of $c(E)$ and $h(j)$. The overall shapes
of them are rather similar; nevertheless, relatively small
differences lead to quite different predictions for the UHECR source
density. The ``shoulders'' of the curves with Dirac-delta
luminosity distributions got smoother for Schechter's distribution.
The scales on the figures are chosen to cover the 95\% confidence regions
(see section \ref{sect_res} for details).

Note, that -- assuming that UHECRs point back to their sources -- our
clustering technique discussed above applies to practically 
any models of UHECR (e.g. neutrinos). One only needs a change in the 
$P(r,E,E_c)$ probability distribution function (e.g. neutrinos penetrate the 
microwave background uninhibited) and use the $h(j)$ and $c(E)$ 
distribution function of the specific model. 

\subsection{Monte-Carlo study of the propagation} \label{sect_monte}

Our Monte-Carlo model of UHECR studies the propagation
of UHECR.
The analysis of
\cite{DMS98} showed that both AGASA and Fly's 
Eye data demonstrated a change of composition, a shift from heavy 
--iron-- at $10^{17}$~eV to light --proton-- at $10^{19}$~eV.
Thus, the chemical composition of UHECRs is most likely
to be dominated by protons. In our analysis we use exclusively protons 
as UHECR particles.
(For suggestions 
that air showers above the GZK cutoff are induced by neutrinos
see \cite{DKDM00}.) 

Using the pion production as the dominant effect of energy loss for
protons at energies $>10^{19}$~eV ref. \cite{BW99} calculated 
$P(r,E,E_c)$, the probability that a proton
created at a given distance (r) with some energy (E) is detected
at earth above some energy threshold ($E_c$). For three
threshold energies the authors of \cite{BW99} gave the approximate 
formula (\ref{bahcall}).

In our Monte-Carlo
approach we determined the propagation of UHECR on an event by event 
basis. Since the inelasticity of Bethe-Heitler
pair production is rather small
($\approx 10^{-3}$) we used a continuous energy loss approximation for
this process. The inelasticity of pion-photoproduction is much higher
($\approx 0.2 -0.5$) in the energy range of interest, thus there are only a
few tens of such interactions during the propagation. Due to the Poisson
statistics of the number of interactions and the spread of the
inelasticity, we will see a spread in the energy spectrum even if the
injected spectrum is mono-energetic.

In our simulation protons propagate in small steps
($10$~kpc), and after each step the energy losses due to pair
production, pion production and the adiabatic expansion are calculated.
During the simulation we keep track of the current energy of the proton
and its total displacement. Thus, one avoids performing new
simulations for different initial energies and distances. The
propagation is completed when the energy of the proton goes below a
given cutoff ($10^{18}$~eV in our case).
For the proton interaction lengths and inelasticities
we used the values of \cite{BS00,AGNM99}. The deflection 
due to magnetic fields is not taken into account, because 
it is small for our typical distances illustrated in Figure 
\ref{f_r}. This fact justifies our assumption that UHECRs point back to
their sources (for a recent Monte-Carlo analysis on deflection
see e.g. \cite{SEMPR00}).

\begin{figure}[!ht]
\begin{center}
\includegraphics[width=8.0cm]{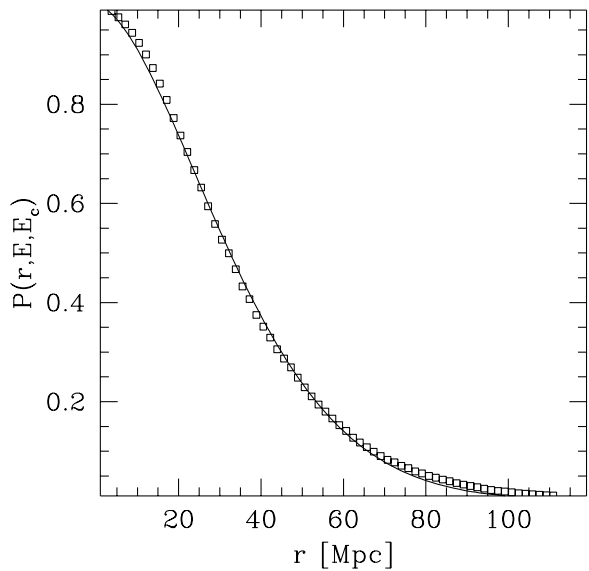}
\caption[a]{\label{fit}
{\sl The direct Monte-Carlo points and the fitted
function $P(r,E,E_c)=\exp\left[ -a\cdot(r/\ {\rm 1Mpc})^b\right]$ for
$E_c=10^{20}$~eV and $E=2\cdot 10^{20}$~eV. The fitted curve
corresponds to $a=0.0019$ and $b=1.695$.
}}
\end{center}\end{figure}

Since it is rather
practical to use the $P(r,E,E_c)$ probability distribution 
function we extended the results of \cite{BW99} by using our Monte-Carlo
technique for UHECR propagation. In order to
cover a much broader energy range than the parametrization
of (\ref{bahcall}) we  used the following type of function
\begin{equation} \label{parametr}
P(r,E,E_c)=\exp\left[ -a\cdot(r/1\ {\rm Mpc})^b\right].
\end{equation}
Figure \ref{fit} demonstrates the reliability of this parametrization. The
direct Monte-Carlo points and the fitted function (eqn. (\ref{parametr}) with
$a=0.0019$ and $b=1.695$) are plotted for
$E_c=10^{20}$eV and $E=2\cdot 10^{20}$eV.

\begin{figure}[p]
\begin{center}
\includegraphics[width=7.3cm]{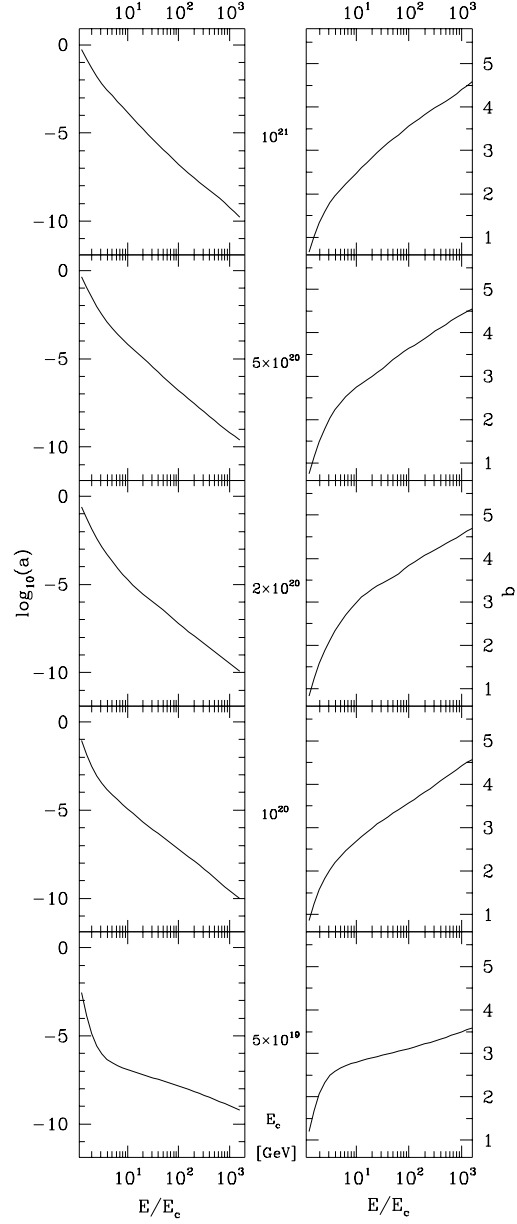}
\caption[a]{\label{gzk}
{ \sl The functions $a(E/E_c)$ --left panel-- and $b(E/E_c)$ 
--right panel-- for the probability distribution function 
$P(r,E,E_c)$ using the parametrization 
$\exp[-a\cdot(r/1\ {\rm Mpc})^b]$ for five different threshold
energies ($5\cdot 10^{19}$~eV, $10^{20}$~eV, $2\cdot 10^{20}$~eV,
$5\cdot 10^{20}$~eV and $10^{21}$~eV).
}}
\end{center}\end{figure}

Figure \ref{gzk} shows the functions $a(E/E_c)$ and $b(E/E_c)$
for a range of three orders of magnitude and for five different
threshold energies. Just using the functions of $a(E/E_c)$ and 
$b(E/E_c)$, thus a parametrization of $P(r,E,E_c)$ one can obtain the 
observed energy spectrum for any injection spectrum without additional
Monte-Carlo simulation.

\subsection{Density of sources}\label{sect_res}

\begin{figure}[!ht]
\begin{center}
\includegraphics[width=8.0cm]{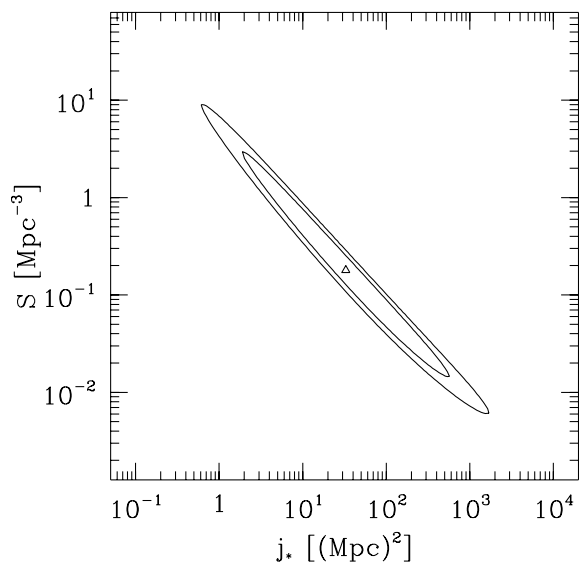}
\caption[a]{\label{ell}
{ \sl The $1\sigma$ (68\%) and $2\sigma$ (95\%) confidence
level regions for $j_*$ and the
source density (14 UHECR with one doublet). 
The most probable value is represented by the triangle. The
upper and lower boundaries of these 
regions give for the source density
$180_{- 165( 174)}^{+2730(8817)}\cdot 10^{-3}$~Mpc$^{-3}$
on the 68\% (95\%)confidence
level.
}}
\end{center}\end{figure}

In order to determine the confidence intervals for the  
source densities we used the frequentist method\cite{PDG}.
We wish to set limits
on S, the source density. Using our Monte-Carlo based
$P(r,E,E_c)$ functions and our analytical technique we
determined $p(N_1,N_2,N_3,...;S;j_*)$, which gives the probability of 
observing $N_1$ singlet, $N_2$ doublet, $N_3$ 
triplet etc. events
if the true value of the density is $S$ and the central value of
luminosity is $j_*$. 
The probability distribution is 
not symmetric and far from being Gaussian. For a given set of 
$\{N_i,i=1,2,...\}$ the above probability distribution as a 
function of $S$ and $j_*$ determines the 68\% and 95\%
confidence level regions in the $S-j_*$ plane.
Figure \ref{ell} shows these regions for our
``favorite'' choice of model ($c(E) \propto E^{-3}$ and
Schechter's luminosity distribution)
and for the present statistics (one doublet out of 14 UHECR events).
The regions are deformed, thin ellipse-like objects in the
$\log(j_*)$ versus $\log(S)$ plane. Since $j_*$
is a completely unknown and independent physical quantity the source density
can be anything between the upper and lower parts of the confidence
level regions. For this model our final answer for the density is
$180_{-165(174)}^{+2730(8817)}\cdot 10^{-3}$~Mpc$^{-3}$,
where the first errors
indicate the 68\%, the second ones in the parenthesis the 95\%
confidence levels, respectively.
The choice of \cite{DTT00} --Dirac-delta like luminosity distribution--
and, for instance, conventional $E^{-2}$ energy distribution
gives much smaller value:
$2.77_{-2.53(2.70)}^{+96.1(916)} 10^{-3}$~Mpc$^{-3}$.
For other choices of $c(E)$
and $h(j)$ see Table \ref{results}. Our results for the Dirac-delta luminosity
distribution are in agreement with
the result of \cite{DTT00} within the error bars. Nevertheless, there is a
very important message.
The confidence level intervals are so large, that on the 95\%
confidence level two orders of magnitude smaller densities than
suggested as a lower bound by \cite{DTT00} are also possible.

As it can be seen there is 
a strong correlation between the luminosity and the
source density. Physically it is easy to understand the picture.
For a smaller source density the luminosities should be larger to give the
same number of events. However, it is not possible to produce the same
multiplicity structure with arbitrary luminosities.
Very small luminosities can not give multiplets at all,
very large luminosities tend to give more than one doublet.

\begin{table}[p]
\begin{center}\begin{tabular}{c|c|c}
$c(E)$ & $h(j)$ & {\bf 14 events 1 doublet} \\
\hline $\propto E^{-2}$ & $\propto \delta$ &
$      2.77 _{-       2.53 (       2.70 )}^{+      96.1  (     916 )}  $\\
\hline $\propto E^{-2}$ & $\propto$ SLF &
$       36.6 _{-        34.3 ( 35.9)}^{+844(4268)}$ \\
\hline $\propto E^{-3}$ & $\propto \delta$ &
$      5.37 _{-       4.98 (       5.25 )}^{+      80.2  (     624 )}  $\\
\hline $\propto E^{-3}$ & $\propto$ SLF &
$       180 _{- 165 ( 174)}^{+2730(8817)}$      \\
\hline $\propto$ decay  & $\propto \delta$ &
$      3.61 _{-       3.30 (       3.51 )}^{+     116    (    1060 )}  $\\
\hline $\propto$ decay  & $\propto$ SLF &
$       40.9 _{-        38.3 ( 40.1)}^{+856(4345)}$ \\
\hline\hline
$c(E)$ & $h(j)$ & {\bf 24 events 1 doublet} \\
\hline $\propto E^{-2}$ & $\propto \delta$ &
$     17.4  _{-      16.0  (      17.0  )}^{+     298    (    2790 )}  $\\
\hline $\propto E^{-2}$ & $\propto$ SLF &
$       200 _{- 169 ( 182)}^{+1230(2428)}$      \\
\hline $\propto E^{-3}$ & $\propto \delta$ &
$     25.0  _{-      22.6  (      24.3  )}^{+     211    (    1690 )}  $\\
\hline $\propto E^{-3}$ & $\propto$ SLF &
$       965 _{- 741 ( 821)}^{+3220(5613)}$      \\
\hline $\propto$ decay  & $\propto \delta$ &
$     20.4  _{-      18.6  (      19.9  )}^{+     358    (    3190 )}  $\\
\hline $\propto$ decay  & $\propto$ SLF &
$       211 _{- 174 ( 190)}^{+1110(2274)}$      \\
\hline\hline
$c(E)$ & $h(j)$ & {\bf 24 events 2 doublets} \\
\hline $\propto E^{-2}$ & $\propto \delta$ &
$      3.19 _{-       2.68 (       2.99 )}^{+      26.4  (     253 )}  $\\
\hline $\propto E^{-2}$ & $\propto$ SLF &
$       41.5 _{-        36.4 ( 40)}^{+424(1514)}$   \\
\hline $\propto E^{-3}$ & $\propto \delta$ &
$      6.42 _{-       5.46 (       6.07 )}^{+      46.2  (     193 )}  $\\
\hline $\propto E^{-3}$ & $\propto$ SLF &
$       208 _{- 182 ( 201)}^{+1970(3858)}$     \\
\hline $\propto$ decay  & $\propto \delta$ &
$      4.18 _{-       3.51 (       3.92 )}^{+      34.5  (     296 )}  $\\
\hline $\propto$ decay  & $\propto$ SLF &
$       45.4 _{-        39.7 ( 43.7)}^{+457(1556)}$ \\
\end{tabular}
\vspace{0.3cm}
\caption[a]{\label{results}
{ \sl The most probable values for the source densities
and their error bars given by the 68\% and 95\% confidence
level regions (the latter in parenthesis). 
The numbers are in units of $10^{-3}$~Mpc$^{-3}$
The three possible energy spectrums 
are given by a distribution proportional to $E^{-2}$, $E^{-3}$, 
or by the decay of a $10^{12}$ GeV particle (denoted by
``decay''). The luminosity distribution can be proportional 
to a Dirac-delta or to Schechter's luminosity function 
(denoted by ``SLF'').
Results are listed for the observed 1 doublet out of 14 events and
for two hypothetical cases (1 doublet out of 24 events and 2 doublets out 
of 24 events). 
}}
\end{center}\end{table}

\begin{figure}[p]
\begin{center}
\includegraphics[width=7.25cm]{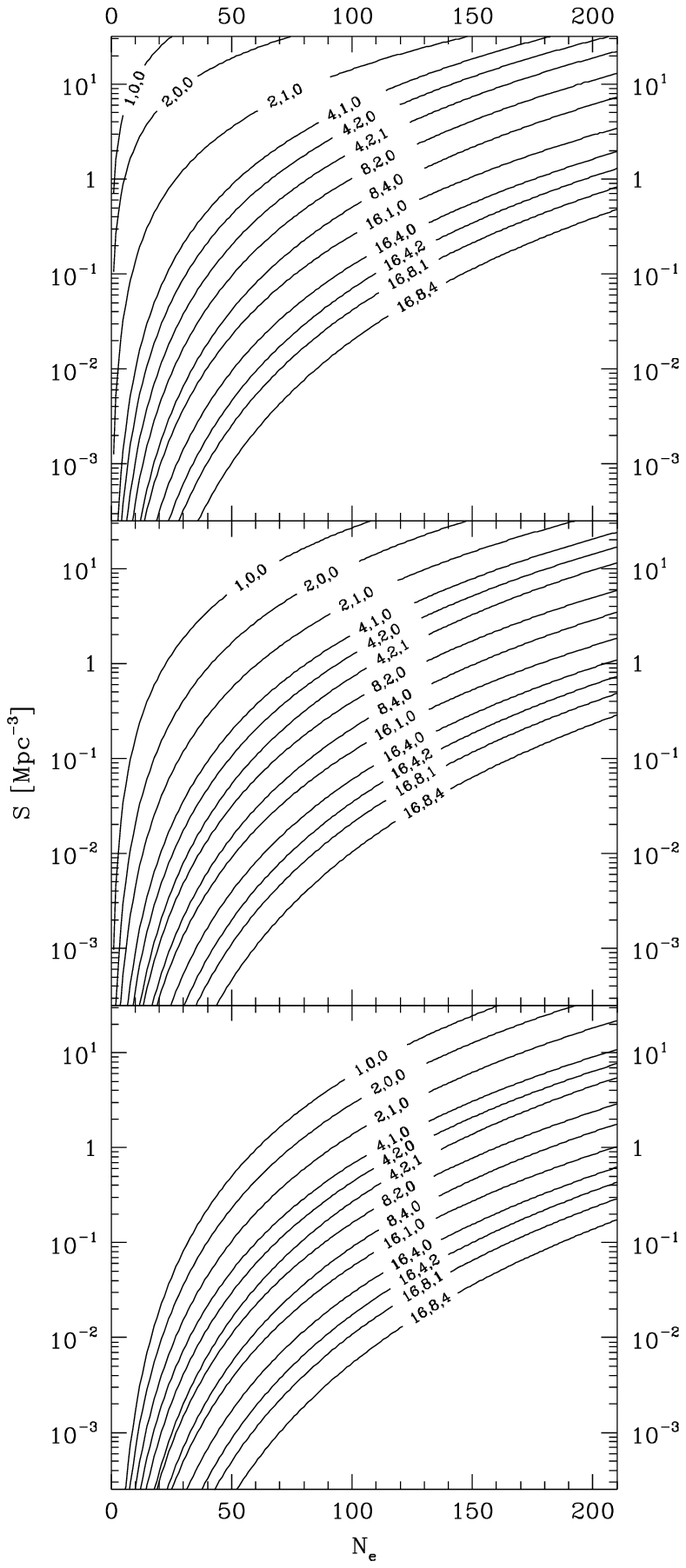}
\caption[a]{\label{center}
{ \sl The most probable values for the density of sources 
as a function
of the total number of events (middle panel). The number of multiplets are
indicated on the individual lines in the form: $N_2,N_3,N_4$, where
$N_2,N_3$ and $N_4$ represent the appropriate values for doublets,
triplets and quartets.
The upper and lower panels correspond to the 84 percentile and
16 percentile lines (upper and lower bounds of the 68\% confidence
intervals), respectively.
}}
\end{center}\end{figure}

The same technique can be applied for any hypothetical experimental
result.
For fixed $\{N_k\}$ 
the above probability function determines the 68\% 
confidence regions in $S$ and $j_*$. 
Using these regions one can
tell the 68\% confidence interval for S. The most probable values 
of the source densities for fixed number of multiplets
are plotted on Figure \ref{center} with the lower and upper bounds. 
The total number of events is shown on the horizontal axis, whereas 
the number of multiplets label the lines. Here again, our ''favorite''
choice of distribution functions was used: $c(E) \propto E^{-3}$ and
$h(j)$ of eqn. (\ref{spread}).

It is of particular interest to analyze in detail the present 
experimental situation having one doublet out of 14 events. 
Since there are some new unpublished events, too, we studied a
hypothetical case of one or two doublets out of 24 events. 
The 68\% and 95\% confidence level results are summarized in Table 
\ref{results} for our three energy and two luminosity distributions. 
It can be seen that Dirac-delta type luminosity distribution
really gives smaller source densities than broad luminosity 
distribution, as it was proven by \cite{DTT00}.
Less pronounced is the effect on the energy distribution of the emitted
UHECRs. The $c(E) \propto E^{-3}$ case gives 
somewhat larger values than the other two choices 
($c(E) \propto E^{-2}$ or given by the decay of a superheavy 
particle). 
The confidence intervals are typically very large, on the 95\%
level they span 4 orders of magnitude. An 
interesting feature of the  results is that ''doubling'' the present
statistics with the same clustering features (in the case studied in
the table this means one new doublet out of 10 new events) reduces
the confidence level intervals by an order of magnitude. The reduction
is far less significant if we add singlet events only. Inspection of
Figure \ref{center}  leads to the conclusion that experiments in
the near future
with approximately 200 UHECR events can tell at least the order of 
magnitude of the source density.

\section{Energy scale of UHECR sources} \label{sect_gut}

An interesting idea suggested by refs.\cite{BKV97,KR98} is that 
superheavy particles (SP) as dark matter could be the source of UHECRs. 
(Metastable relic SPs were proposed 
\cite{ELN90} before the observation of UHECRs beyond 
the GZK cutoff.) In 
\cite{KR98} extragalactic SPs were studied.
Ref. \cite{BKV97} made a crucial observation and analyzed the
decay of SPs concentrated in the 
halo of our galaxy. They used  the modified leading logarithmic
approximation (MLLA) \cite{MLLA} for ordinary QCD and for supersymmetric QCD
\cite{BK98}. 
A good agreement of the extragalactic spectrum with observations was
noticed in \cite{BBV}.
Supersymmetric QCD is treated as the strong regime of MSSM.
To describe the decay spectrum more 
accurately HERWIG Monte-Carlo was used in QCD \cite{BS98} and discussed
in supersymmetric QCD \cite{S00,Rubin}, resulting in 
$m_X \approx 10^{12}$ GeV and $\approx 10^{13}$ GeV for the SP mass in  
SM and in MSSM, respectively. 

SPs are very 
efficiently produced by the various mechanisms at post inflatory 
epochs \cite{B00}. Our analysis of SP decay 
covers a much broader class of possible sources. 
Several non-conventional UHECR sources
(e.g. extragalactic long ordinary strings \cite{VAH98} or galactic
vortons \cite{MS98}, monopole-antimonopole pairs connected by strings
\cite{BO99}) produce the same UHECR spectra as decaying SPs. 

In my thesis I study the scenario that the UHECRs are coming
from decaying SPs and I determine the mass $m_X$ of this $X$ 
particle by a detailed analysis of the observed UHECR spectrum. 
I discuss both possibilities that the UHECR protons are 
produced in the halo of our galaxy and that they are of extragalactic
origin and their propagation is affected by CMBR. We did not investigate
how can they be of halo or extragalactic origin, we just analyzed their
effect on the observed
spectrum instead.
We assumed that the SP decays into two quarks (other decay modes 
would increase $m_X$ in our conclusion). 
After hadronization these quarks yield protons. The result is characterized 
by the fragmentation function (FF) $D(x,Q^2)$ which gives the
number of produced 
protons with momentum fraction $x$ at energy scale $Q$.
For the proton's FF at present accelerator
energies we use ref. \cite{BKK95}. We evolve
the FFs in ordinary \cite{DGLAP} and 
in supersymmetric \cite{JL83} QCD to the energies of the 
SPs. This result can be combined with the
prediction of the MLLA technique 
, which gives 
the initial spectrum of UHECRs at the energy $m_X$. 
Altogether we studied four different models:
halo-SM, halo-MSSM, EG-SM and EG-MSSM. 


\subsection{Decay and fragmentation of heavy particles}

As in the previous sections we again assumed that
UHECRs are
dominated by protons and in our analysis we used them exclusively.

The FF of the proton can be determined from 
present experiments \cite{BKK95}. 
The FFs 
at $Q_0$ energy scale are
$D_i(x,Q_0^2)$,
where $i$ represents the different partons (quark/squark or gluon/gluino).
The FFs can not be determined
in perturbative QCD. However, their evolution in $Q^2$ 
is governed by the Dokshitzer-Gribov-Lipatov-Altarelli-Parisi (DGLAP)
equations \cite{DGLAP}:
\begin{equation} 
{\partial D_i(x,Q^2) \over \partial \ln Q^2}= 
\frac{\alpha_s(Q^2)}{2\pi}
\sum_j \int_x^1 {dz \over z} P_{ji}(z,\alpha_s(Q^2))D_j(\frac{x}{z},Q^2),
\end{equation} 
One can interpret
$P_{ji}(z)$, the splitting function, as the probability density that
a parton $i$ produces a parton $j$ 
with momentum fraction $z$.

The direct solution of the DGLAP equations is rather difficult. We can
introduce the moments of the FFs and splitting functions:
\bee
M_i(n)=\int_0^1{x^{n-1}D_i(x)dx}
\ee
\bee
A_{ji}(n)=\int_0^1{x^{n-1}P_{ji}(x)dx}
\ee
In terms of these moments the DGLAP equations have a simple form. Using the
leading order expression for the running coupling constant $\alpha_s(Q^2)$
one gets:
\bee \label{eq_mom}
\frac{\partial M_i(n,t)}{\partial t}=\frac{1}{\beta_0 t}
\sum_j{A_{ji}(n)M_j(n,t)}
\ee
where $t=\ln (Q^2/\Lambda^2)$ and $\beta_0=\frac{33-2N_f}{6}$ for QCD
and $\beta_0=\frac{27-3N_f}{6}$ for SUSY-QCD evolution.
These linear differential equations are easy to solve.

We solved the DGLAP equations using this method numerically with the 
conventional QCD splitting functions (for the SM scenarios) 
and with the supersymmetric 
ones (for the MSSM scenarios) \cite{JL83}. We started from the FFs of ref. \cite{BKK95}.
For the top and the MSSM partons at their threshold energies
we used the FFs of ref. \cite{Rubin}. While solving the DGLAP equations each 
parton was included at its own threshold energy. As the energy increases,
the number of flavors involved increases and so $\beta_0$ decreases. Thus
$\Lambda$
should be adjusted in order to make the right hand side of (\ref{eq_mom})
continuous.

\begin{figure}[!ht]
\begin{center}
\includegraphics[width=8.0cm]{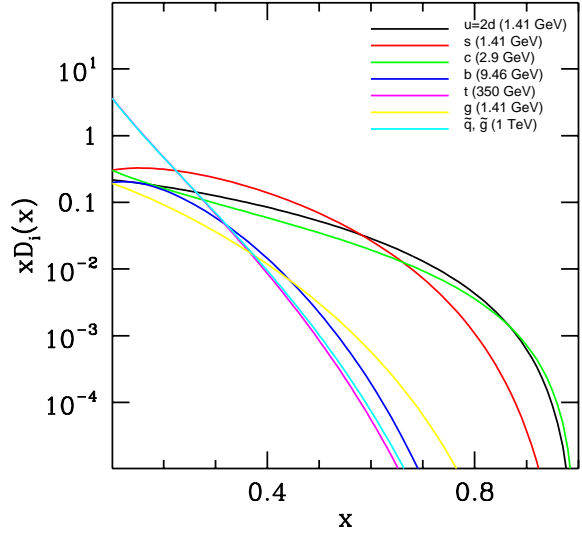}
\caption[a]{\label{fig_frag0}
{\sl The fragmentation functions of Table \ref{frag_tab} for the different
partons.
}}
\end{center}\end{figure}

We checked that our
final result on $m_X$ is insensitive to the choices of the top and MSSM parton
FFs. The main difference between the SM and MSSM cases came from
the different $\beta$ functions.
Table \ref{frag_tab} and Figure \ref{fig_frag0} 
shows all the initial FFs we used
at 
different energy scales indicated in the second column of the table.

\begin{table}[!ht]
\begin{center}\begin{tabular}{l|l|l|l|l}
flavor&$Q_0$~(GeV)&$N$&$\alpha$&$\beta$\\
\hline
$u=2d$ &1.41&0.402&-0.860&2.80\\
\hline
$s$    &1.41&4.08&-0.0974&4.99\\
\hline
$c$    &2.9&0.111&-1.54&2.21\\
\hline
$b$    &9.46&40.1&0.742&12.4\\
\hline
$t$    &350&1.11&-2.05&11.4\\
\hline
$g$    &1.41&0.740&-0.770&7.69\\
\hline
$\tilde{q}_i, \tilde{g}$&1000&0.82&-2.15&10.8\\
\end{tabular}
\vspace{0.3cm}
\caption[a]{\label{frag_tab}
{\sl The fragmentation functions of the different partons using the 
parametrization $D(x)=Nx^\alpha (1-x)^\beta$ at different energy scales
(second column).
}}
\end{center}\end{table}

The fragmentation functions after beeing evolved to high scales
can be well parametrized as:
\bee
D(x)=N_1x^{\alpha_1} (1-x)^{\beta_1}+N_2x^{\alpha_2} (1-x)^{\beta_2}.
\ee
Table \ref{tab_frag1} gives the fragmentation functions averaged over
the quark flavors for the energy range $10^{12}$~eV-$10^{17}$~eV in this
parametrization.

\begin{table}[!ht]
\begin{center}\begin{tabular}{l|l|l|l|l|l|l}
log(Q/eV)&$N_1$&$\alpha_1$&$\beta_1$&$N_2$&$\alpha_2$&$\beta_2$\\
\hline\hline
SM \\
\hline
12.&	0.0186& -1.28& 3.56& 0.199& -1.70& 5.26\\
\hline
13.&	0.0194& -1.30& 3.66& 0.186& -1.71& 5.36\\
\hline
14.&	0.0207& -1.31& 3.76& 0.175& -1.72& 5.47\\
\hline
15.&	0.0215& -1.32& 3.84& 0.165& -1.73& 5.56\\
\hline
16.&	0.0225& -1.35& 3.93& 0.157& -1.74& 5.66\\
\hline
17.&	0.0232& -1.35& 4.01& 0.150& -1.75& 5.75\\
\hline\hline
MSSM \\
\hline
12.&	0.026& -1.37&  4.23& 0.106 & -1.77& 6.03 \\
\hline
13.&	0.027& -1.39&  4.41& 0.0941& -1.79& 6.30 \\
\hline
14.&	0.028& -1.22&  4.57& 0.0890& -1.80& 6.51 \\
\hline
15.&	0.028& -0.735& 4.70& 0.0855& -1.80& 6.48 \\
\hline
16.&	0.029& -0.421& 4.85& 0.0785& -1.81& 6.55 \\
\hline
17.&	0.030& -0.441& 5.00& 0.0724& -1.82& 6.76 \\
\end{tabular}
\vspace{0.3cm}
\caption[a]{\label{tab_frag1}
{\sl The fragmentation functions of the proton
averaged over the quark flavors for high
energies in the SM and MSSM using the parametrization 
$D(x)=N_1x^{\alpha_1} (1-x)^{\beta_1}+N_2x^{\alpha_2} (1-x)^{\beta_2}$
}}
\end{center}\end{table}

The FFs obtained this way are not accurate for very small $x$ values,
since even the original FFs are not well known in this region.

At small $x$ values multiple soft gluon emission can be 
described by the MLLA \cite{MLLA}.
This gives the shape of the total hadronic FF  for soft particles (not distinguishing 
individual hadronic species)
\bee
xF(x,Q^2)\propto \exp \left[ -\ln(x/x_m)^2/(2\sigma^2)\right],
\ee
which is peaked at $x_m = \sqrt{\Lambda/Q}$ with 
$2\sigma^2= A\ln^{3/2}(Q/\Lambda)$.
According to \cite{BK98}
the values of $A$ are 
$\sqrt{7/3}/6$ and $1/6$ for SM and MSSM, respectively.
The MLLA describes the observed hadroproduction quite accurately
in the small $x$ region \cite{DELPHI}.
For large values of $x$ the MLLA should not be
used.

\begin{figure}[!ht]
\begin{center}
\includegraphics[width=8.0cm]{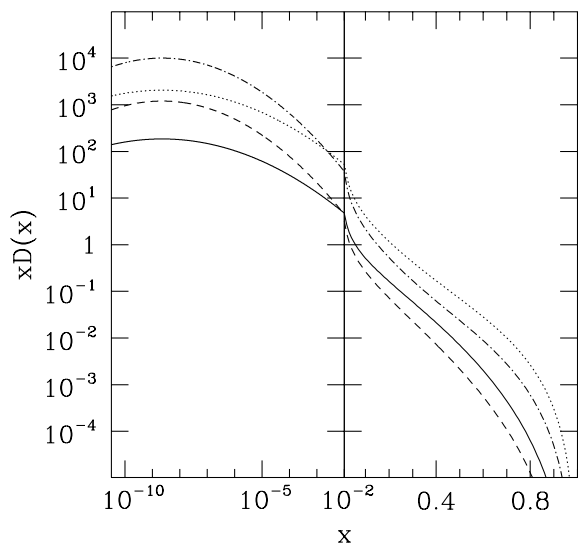}
\caption[a]{\label{fragmentation}
{\sl The FFs averaged over the quark flavors
at $Q=10^{16}$ GeV for proton/pion in SM (solid/dotted line)  and in MSSM 
(dashed/dashed-dotted line)
in the relevant $x$ region. To show both the small and large $x$
behavior we change from logarithmic scale to linear at $x=0.01$.
}}
\end{center}\end{figure}

We smoothly connected the solution for the FF
obtained by the DGLAP equations 
and the MLLA result at a given $x_c$ value.
Our final result on $m_X$
is rather insensitive to the choice of $x_c$, the uncertainty is included
in our error estimate.
The SP decay also produces a huge number of pions. The total number
of produced pions is essential since they decay to photons which loose
most of their energy during propagation and give a contribution to the low
energy photon spectrum.
Thus we also determined the FF of the pion.
Figure \ref{fragmentation} shows the FF for the proton and 
pion at $Q=10^{16}$~GeV in  SM and MSSM.

\subsection{Comparison of the predicted and the observed spectra}

UHECR protons produced in the halo of our galaxy can propagate
practically unaffected and the production spectrum should be
compared with the observations. 

Particles of extragalactic origin
and energies above $\approx 5\cdot 10^{19}$ eV loose a large fraction 
of their energies due to interactions with CMBR 
\cite{GZK66}. This effect can be quantitatively described by the
function $P(r,E,E_c)$ introduced and calculated in section \ref{sect_monte}.
The original UHECR spectrum 
is changed at least by two different ways: (a) there should be a 
steepening due to the GZK effect; (b) particles loosing their
energy are accumulated just before the cutoff and produce a bump. 
We studied the observed spectrum by
assuming a uniform source distribution for UHECRs.


Our analysis includes the published and the unpublished
UHECR data 
of \cite{AGASA,FLY,HAVERAH,HIRES}. Due to normalization difficulties
we did not use the Yakutsk \cite{YAKUTSK} results. 
We also performed the analysis using the AGASA data only and found the 
same value (well within the error bars) for $m_X$.
Since the decay of SPs results 
in  a non-negligible flux for lower energies $\log (E_{min}/\mbox{eV})=18.5$ 
was used as a lower end for the UHECR spectrum. Our results are
insensitive to the definition of the upper end (the flux is
extremely small there) for which we chose $\log (E_{max}/\mbox{eV})=26$.
As it is usual we divided each logarithmic unit into ten bins. The
integrated flux
gives the total number of events in a bin. The uncertainties of the 
measured energies are about 30\% which is one bin. Using a Monte-Carlo method
we included this uncertainty in the final error estimates.
The predicted number of events in a bin is given by
\begin{equation}\label{flux}
N(i)=\int_{E_i}^{E^{i+1}}\left[A \cdot E^{-3.16}+B\cdot j(E,m_X)\right],
\end{equation}
where $E_i$ is the lower bound of the $i$-th energy bin. The first 
term describes the data below $10^{19}$~eV according to
\cite{AGASA}, where the SP decay gives negligible contribution.
The second one corresponds to the spectrum of the decaying
SPs. A and B are normalization factors.

\begin{figure}[!ht]
\begin{center}
\includegraphics[width=8.0cm]{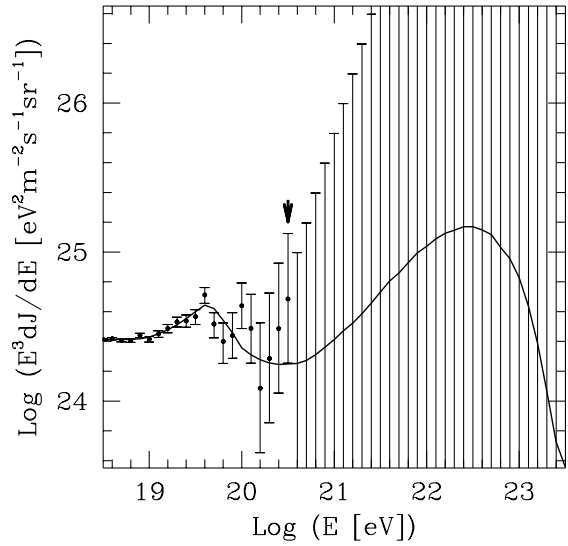}
\caption[a]{\label{spect}
{\sl The available UHECR data with their error bars
and the best fit from a decaying SP using the EG-MSSM scenario.
Note that there are no events above $3 \times 10^{20}$~eV 
(shown by an arrow). 
Nevertheless the experiments are sensitive even in this region. Zero event
does not mean zero flux, but a well defined upper bound for the flux 
(given by the Poisson distribution).
Therefore the experimental value of the
integrated flux is in the ''hatched'' region with 68\% confidence level. 
(''hatching'' is a set of individual 
error bars; though most of them are too large to be depicted in full) 
Clearly, the error bars are large enough to be consistent with the SP decay.
}}
\end{center}\end{figure}

\begin{figure}[!ht]
\begin{center}
\includegraphics[width=8.0cm]{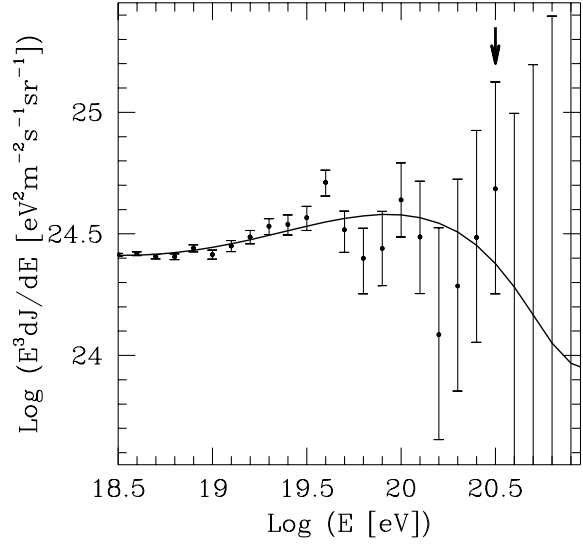}
\caption[a]{\label{spect_halo}
{\sl The available UHECR data with their error bars
and the best fit using the halo-MSSM scenario.
}}
\end{center}\end{figure}

The expectation value for the number of events in a bin is given
by eqn. (\ref{flux}) and it is Poisson distributed. To
determine the most probable $m_X$ value we used the maximum-likelihood
method by minimizing the $\chi^2(A,B,m_X)$ for
Poisson distributed data \cite{PDG}
\begin{equation} \label{chi}
\chi^2=\sum_{i=18.5}^{26.0}
2\left[ N(i)-N_o(i)+N_o(i)\ln\left( N_o(i)/N(i)\right) \right],
\end{equation}
where $N_o(i)$ is the total number of observed events in the $i$-th
bin. In our fitting procedure we had three parameters: $A,B$ and $m_X$.
The minimum of the $\chi^2(A,B,m_X)$ function is $\chi^2_{min}$ 
at $m_{X min}$ which is 
the most probable value for the mass, whereas
$\chi^2(A',B',m_X)\equiv \chi^2_o(m_X)=\chi^2_{min}+1$
gives the one-sigma (68\%) confidence interval for $m_X$. 
Here $A',B'$ are defined in 
such a way that the $\chi^2(A,B,m_X)$ function is minimized in $A$ 
and $B$ at fixed $m_X$.
Figure \ref{spect} shows the measured UHECR spectrum and the best fit in the
EG-MSSM scenario.
The first bump of the fit represents particles produced at
high energies and accumulated just above the GZK cutoff due to their energy
losses. The bump at higher energy is a remnant of $m_X$. In the halo
models there is no GZK bump (Figure \ref{spect_halo}), so the relatively large 
$x$ part of the FF moves
to the bump around $5\times 10^{19}$~GeV resulting in a much smaller $m_X$ than
in the extragalactic case. An interesting feature of the GZK effect is that the shape of
the produced GZK bump is rather insensitive to the injected spectrum so the
dependence of $\chi^2$ on the choice of the FF is small.
The experimental data is far more accurately described by the 
GZK effect (dominant feature of the extragalactic fit) than by the FF itself (dominant for 
halo scenarios).

\subsection{Predictions for $m_X$}

\begin{figure}[!htb]
\begin{center}
\includegraphics[width=8.0cm]{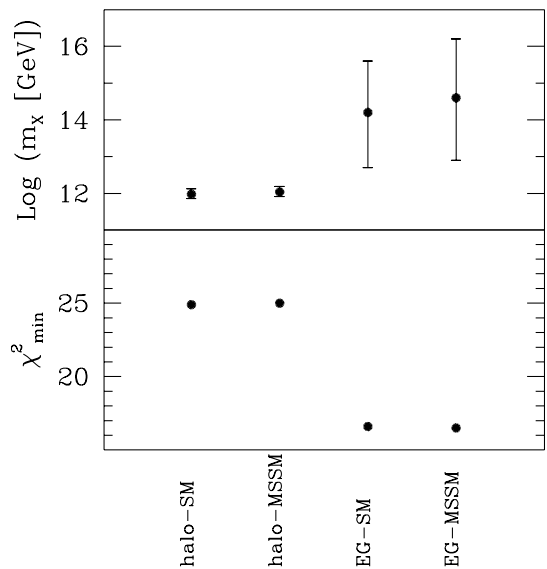}
\caption[a]{\label{result}
{\sl The most probable values for the mass of the decaying
ultra heavy dark matter with their error bars and the
total $\chi^2$ values. Note that 21 bins contain nonzero number of events
and eqn.(\ref{flux}) has 3 free parameters.
}}
\end{center}\end{figure}

To determine the most probable value for the mass of the 
SP we studied four scenarios. Table \ref{tab_result} and 
Figure \ref{result} contain
the $\chi^2_{min}$ values and the most
probable masses with their errors for these scenarios.

The UHECR data favors the EG-MSSM scenario. The predicted mass is
$10^b$~GeV, where $b=14.6_{-1.7}^{+1.6}$.
The
goodnesses of the fits for the halo models are far worse. 
The SM and MSSM cases do not differ significantly. 
The most important message is that the masses of the best fits 
(extragalactic cases)
are compatible within the
error bars with the MSSM gauge coupling unification GUT scale \cite{ABF92}.
\begin{table}[!ht]
\begin{center}\begin{tabular}{l|l|l}
scenario&$\chi^2$&$\log_{10}(m_X/\mbox{GeV})$ \\
\hline
halo-SM	&	24.9&	$11.98^{+.15}_{-.12}$\\
\hline
halo-MSSM&	25.0&	$12.04^{+.15}_{-.12}$ \\
\hline
EG-SM	&	16.6&	$14.2^{+1.4}_{-1.5}$ \\
\hline
EG-MSSM	&	16.5&    $14.6^{+1.6}_{-1.7}$\\
\end{tabular}
\vspace{0.3cm}
\caption[a]{\label{tab_result}
{\sl The $\chi^2$ and $m_X$ values for the four scenarios.
}}
\end{center}\end{table}

The SP decay will also produce a huge number of pions which will decay into 
photons. Our spectrum contains 94\% of pions and 6\% of protons.
This $\pi /p$ ratio is in agreement with the calculations of
\cite{Sigl99,BS00} which showed that
for different classes of models with $m_X \lsim 10^{16}$~GeV , which is the upper boundary
of our confidence intervals, the generated
gamma spectrum is still consistent with the observational constraints. 

In the near future the UHECR statistics will probably be increased by an
order of magnitude \cite{PAUG}. Performing our analysis for such a
hypothetical statistics the uncertainty of $m_X$ was found to
be reduced by two orders of magnitude.

Since the decay time of the SPs should be at least
the age of the universe it might happen that such SPs
overclose the universe. Due to the large mass of the SPs a single decay 
results in a large number of UHECRs, thus a relatively small 
number of SPs can describe the observations. We calculated the  
minimum density required for the best-fit spectrum in each scenario
and it was more than ten orders of magnitude smaller than the critical one.

\chapter{The Poor Man's Supercomputer} \label{PMS}

In this chapter I briefly describe the hardware and software architecture of
the Poor Man's Supercomputer (PMS) built at the E\"otv\"os University \cite{PMS00}.
This supercomputer was used to perform the lattice simulations of
chapter \ref{MSSM} as well as the Monte-Carlo analysis of chapter \ref{UHECR}.


Our purpose was to build a high performance supercomputer from PC elements.
We use PC's for two reasons. They have excellent cost/performance ratios
\cite{pricewatch} and
can easily be upgraded when faster motherboards and CPUs will be available.

The PMS project
started in 1998, and the machine has been ready for physical calculations since
the spring of 2000.
Our first PMS machine (PMS1) consists of 32 PC's arranged in a three-dimensional 
$2 \times 4 \times4$ mesh. Each node has two special
communication cards providing fast communication through flat cables
to the six neighbours.
This gives a much better performance than one Ethernet Token Ring.

The following sections describe the hardware and software architectures
of PMS. First a short overview of the machine is given and then the 
hardware and the software aspects are described
in more details. Some performance results are also presented.

\section{Overview}
The nodes in PMS are based on PC components. Our first PMS machine (PMS1)
contains 32 rack-mounted nodes.
Each node is powered by its own standard PC power supply located at the
bottom of the rack. 

Each node in PMS is an almost complete PC.
In PMS1 the current configuration of a node consists of a 
100MHz motherboard (SOYO SY-5EHM), 
a single 450MHz  AMD K6-II processor,
128MB (7ns) SDRAM, 
10Mb Ethernet card, 
and a hard disk of capacity 2.1 Gbyte.
The nominal speed of each node is 225 Mflops, since floating point
operations of the AMD processor require two clock cycles.

PMS uses special hardware  for communication (PMS CH)
to make high speed parallel
calculations possible.   
The basic idea behind the hardware is that the PMS CH  provides each node
 with direct connection 
to its nearest neighbors. Our first implementation of the PMS CH includes two plug-in ISA cards, 
the PMS CPU card and the PMS Relay card. 
The PMS CH handles both polled port (PP) IO operations and direct memory 
access (DMA) between two selected nodes. However, DMA is not used at present.

Programming  the PMS CH is a fairly simple task. It is currently done under 
 Linux. 
All low-level device drivers are written
in C and the programmer may use all kinds of commercial, share-ware or free-ware
compilers. Using the communication drivers requires only the knowledge of a 
few functions.

\begin{figure}[!ht]
  \centerline{\includegraphics[width=8.0cm]{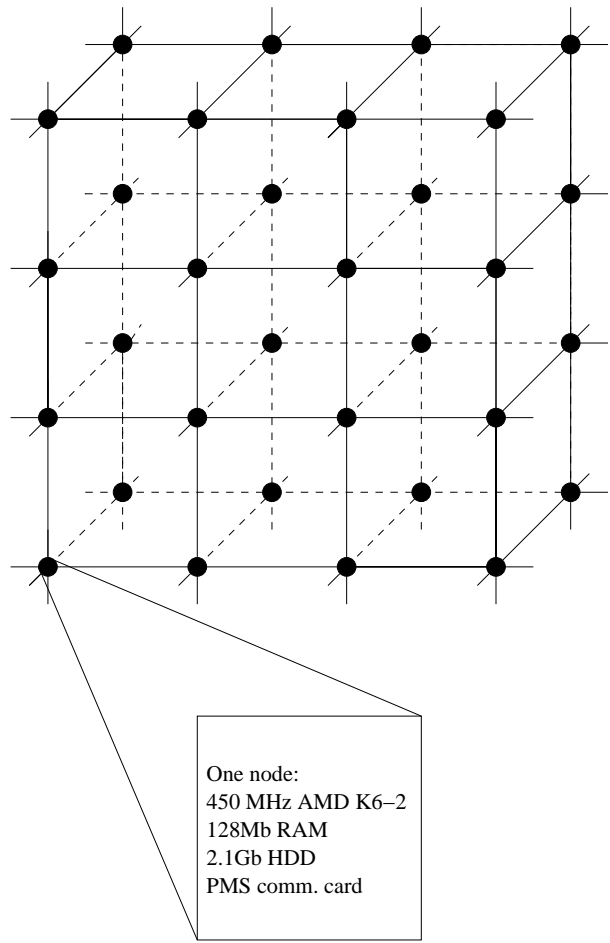}}
  \caption[a]{{\sl The PMS cluster}} \label{pms_grid}
\end{figure}

The nodes are arranged in a $2 \times 4 \times4$  mesh as shown in Figure 
\ref{pms_grid}.
In each node both the  PMS CPU and the PMS Relay cards are installed
providing fast communication to the six nearest neighbors.
At the boundaries periodic boundary conditions are realized as indicated in 
Figure \ref{pms_grid}, where the links at the boundaries correspond to the ones on the other sides.
This determines the hardware architecture of the machine, which is
similar to that of the APE machines \cite{APE}.

 Debian Linux 2.1 is installed on each node. After turning the power on
each node boots from its own hard-disk. All nodes can be accessed through the
Ethernet Token Ring. There is a main computer that controls the whole cluster.
A tiny job-management system was written to copy the executable
program code
and the appropriate data to and from the nodes, to execute the programs and
to collect the results.
In principle, the Ethernet Token Ring could be used for data transfer between 
the nodes during simulations. However, 
this turned out to be too slow in most cases. 
One major reason for this is
that any data transfer between two machines makes the whole network busy.
Since building up the Ethernet connection (protocol overhead) is quite 
slow even for two
computers, the Ethernet Token Ring is not satisfactory. 

The special communication cards --described in more detail in the next
section-- provide faster communication between adjacent nodes. However, this
makes the machine applicable only to local problems, where
communication between the neighboring nodes is necessary. In PMS1 
the speed of 
communication through these cards --limited essentially by the ISA bus speed-- 
is about 2 MB/s. The measured value in real applications  
, i.e. the speed of a link for a given direction is
essentially the same: 2.2MB/s.  
The time needed to build up our 
communication is negligible for package sizes over 1kB.  
Furthermore, 16 pairs of machines can
simultaneously communicate. 
The system is scalable.
One can build machines with larger number of nodes. The
total inter-node communication performance is proportional to the
number of nodes.

\section[Description of the PMS 
Communication Hardware (CH)]{Description of the PMS Communication\\
Hardware (CH) }

There are two communication cards in each machine. The CPU card contains
the main
circuits needed for transmitting data, while the Relay card contains the
connectors for the flat cables connecting the adjacent nodes and some
additional circuits.
\begin{figure}[!ht]
  \centerline{\includegraphics[width=8.0cm]{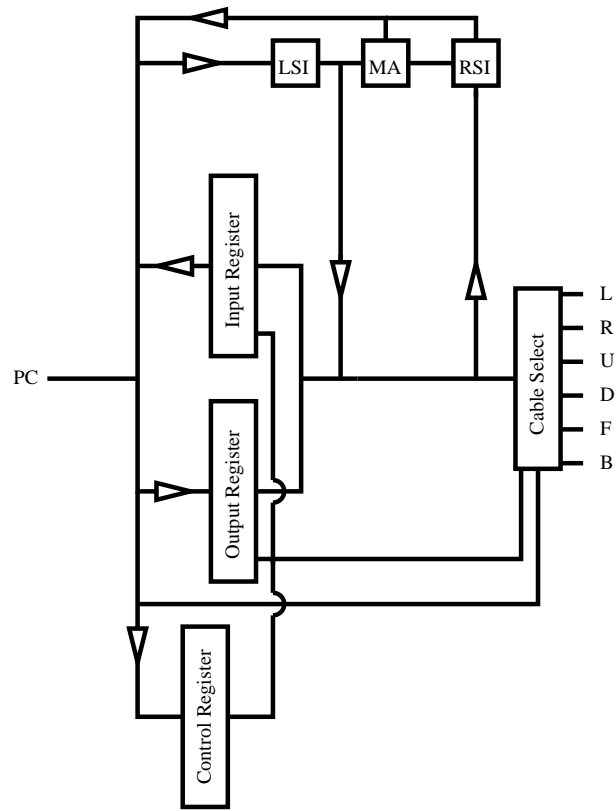}}
  \caption[a]{{\sl The PMS communication card}} \label{card_scem}
\end{figure}
The block diagram of the cards is shown in Figure \ref{card_scem}. 
The circuits of both
cards are included. There are two 16-bit buffers,
the output buffer and the input buffer, which accept the data coming from the
computer and from one of the adjacent nodes, respectively. The Control
Register
is used --among other things-- to clear the buffers and set the node to
either
sender or receiver state.
The Cable Select
circuit selects which direction the data is sent to or received from. The six
directions are labeled as left (L), right (R), up (U), down (D), front (F),
back (B).
If the same physical cable is selected by two adjacent nodes, one of them being
set
as sender
and the other as receiver, a physical connection is established and the
content of the output register of the sender is
immediately transferred to the receiver's input register.
The Local State Indicator (LSI) and the Remote State Indicator (RSI) are two
registers
to indicate the states of the nodes. There are 12 LSI and 12 RSI lines. They
correspond
to sending to and receiving from the six directions.
Each node can indicate its request for sending or receiving through the LSI
lines.
The RSI lines are identical to the six neighbors' LSI lines.
If there is a match between RSI and LSI signals (i.e. a send and a
corresponding receive request coincide) then the Match Any (MA) bit is set
and an IRQ
is generated on the PC bus. The interrupt is generated on both nodes at the
same time, so
the interrupt handlers on both machines can safely start transferring data
without any extra synchronization. The data can be transferred either 
by DMA  (not used at present) or by consecutive I/O operations.

The Cable Select circuit, the LSI, RSI and MA registers together with the
flat
cable connectors are located on the relay card, while all other circuits are
located on the CPU card.

\section{Software}

As  mentioned above, the whole cluster is connected via an Ethernet
Token Ring. Each node has a special hostname that corresponds
to its location in the cluster. 
The hostnames
are s000,s001,s002,s003,s010 $\dots$ s133. The numbers in the 
hostname correspond to the coordinates of the node in the
three dimensional mesh.
When it is necessary for a node to identify itself (e.g. write/read priorities
during inter-node communication, see below) the
file '/etc/hostname' is used. The Ethernet Token Ring is used only 
for job management, i.e. to distribute
and collect data to and from the nodes. It is not used for 
inter-node communication during simulations. This is achieved 
by the special hardware described in the previous section.

In order to take advantage of the fast communication from applications
(e.g. high level C, C++ or Fortran code),
a low-level Linux kernel driver has been developed to access all the registers
of the communication card. From the user level cards can simply be reached by
reading or writing the device files '/dev/pms0, /dev/pms1 ... /dev/pms5'.
The six device files correspond to the six directions. A write operation
to one of these device files will         
transmit data to the corresponding direction, and reading from these device 
files reads
out previously transferred data. All the necessary input/output operations for
transferring data blocks are performed by the device driver. Notice the
important feature that this can be reached from any high level 
C, C++, Fortran code for which compilers are available.

The main structure of the device driver is similar to that of the card.
There are six read buffers, one for each direction in the main memory of the
machine and there is one write buffer. The data are always written to the write
buffer and read out from one of the read buffers.

The driver has two main parts. The first part is accessed from applications when
the user writes or reads any of the device files '/dev/pms*'. The other part is
the interrupt handler where the real data transfer takes place.

Whenever data are written to one of the device files, all the driver does is to
copy the written data to the write buffer and set the corresponding LSI send
signal to
indicate that a data send is requested. If the buffer is already full,
an error byte is returned to the application. Reading from the device files is
similar: if there are data in the corresponding read buffer, they are sent to the
application and the LSI receive line is set, since the node is ready to receive
new data. If the read buffer is empty, an 'End Of File' byte is returned to the
application.

When there is a coincidence between corresponding LSI and RSI signals,
an interrupt is generated by the card, which invokes the driver's
interrupt handler. It is the task of the handler to transfer data from the
sender to the receiver.
The interrupt handlers on the two communicating nodes start almost at
the same time. The difference may only be a few clock cycles. There is,
however, a need for synchronization. If the machines are ready to send
or receive the first byte they indicate it with their LSI lines. Notice 
that this will not cause an extra interrupt since interrupts are
disabled within the interrupt handler.
The sender first transmits the size of the package that will
follow in 16-bit words.
In the present version this is a 16-bit value, so the maximum size of a
package that can be transferred is 128 kbytes. Then the given
number of 16-bit values follow. Each word from the sender's write buffer
is copied to the receiver's read buffer. Finally, a 32-bit checksum is sent.
The receiver computes its own checksum and if it does not match the
received checksum, it is indicated to the sender and the whole transfer
is repeated. The final step in the interrupt handler is to clear the 
LSI lines of both nodes. On the one hand this indicates for the sender that the data
have been transferred and
the write buffer is empty again. On the other hand this tells the
receiver that the corresponding read buffer is full, so no new data can
arrive unless the buffer is emptied.

The buffer sizes are set in the driver to constant values. From the
previous paragraph it is clear that the maximum reasonable buffer size
is 128 kbytes. In order to save memory, while allowing large packages
at the same time the buffer size is set to only 64 kbytes at present.

The driver makes application programming quite easy. Communication can
be achieved by accessing the above mentioned device files. 
However, some C functions
have also been written to make writing applications even
simpler. These functions are the following:

{\it pms\_open} is used to initialize the card. It clears all
buffers, sets the LSI receive lines, clears the LSI send lines and
enables interrupts. The node thus becomes ready to receive data from any of
the neighbors.

{\it pms\_close} is used to close the card. All LSI lines
are cleared and interrupts are disabled. No further
communication may take place after this function call.

{\it pms\_send, pms\_recv}  are used to send and
receive data. Their parameters are the direction, the number of bytes
to send, and a pointer to the beginning of the data. On success they
return a positive value, otherwise a negative one. If there is no data
in the read buffer, {\it pms\_recv} returns 0.

{\it pms\_send\_receive} is a commonly used combination of {\it pms\_send}
and {\it pms\_recv}. It sends data to the specified direction and receives
data from the opposite direction. The order of send and receive depends on
the parity of the node as discussed later.

{\it pms\_collect, pms\_average} are global operations that simply collect or
take the average of data from all nodes in one direction.

The driver does not take care of any priority problems. It is possible
to write applications that will not work since all nodes are waiting
for data while none of them is sending anything. This is often the case
when the same code is running on all the nodes without any priority
check. There is a simple solution to these kind of problems. The
parity of the node is simply the
parity of the sum of the three digits in its hostname. Each time when
communication is performed, even nodes send data first and receive
afterwards, while odd nodes receive first and send their data afterwards.
This simple method is used in the {\it pms\_send\_receive} function.
Applications that use only this function to transfer data should not worry
about priority problems.

\section{Performance}

The lattice simulations of chapter \ref{MSSM} were carried out
on our parallel
computer PMS1. We also measured the performance of the machine with
a much simpler pure SU(3) gauge theory.
The most CPU time consuming parts, manipulation with
$2\times 2$ and $3\times 3$ complex
matrices, were written in assembly language.
This increased the speed of the codes by about a factor 
of two.

We obtained similar results for the speed of the
code and for the communication between nodes in the two cases. 
The MSSM results for communication are actually somewhat better.
The reason
for that is quite simple. The number of variables in the MSSM
is  larger by a factor of two than in pure SU(3) gauge theory; 
however, the number of floating point operations 
needed for a full update is more than an order of magnitude larger.
Thus, for the same lattice size the time needed to
transfer the surface variables --done by the
communication cards-- compared to the update time
is smaller for MSSM than for the pure SU(3) theory. 
We estimated that a  similar or a bit less speed than in SU(3) gauge theory  
should be observed for fermionic systems  
with multiboson \cite{multiboson} algorithms.

For small lattice sizes the most economical way to use our
32 PC cluster is to put independent lattices on the different
nodes. The maximum lattice size in the SU(3) theory 
for 128 MB memory is $\sim 20^4$,
or for finite temperature systems $6 \cdot 32^3$.
One thermalizes such a system on a single node, then distributes
the configuration to the other nodes and continues the
updating on all 32 nodes.
We measured the sustained performance of the cluster in this
case, which gave  $32 \times 152$Mflops=4.9Gflops. This
152Mflops/node performance means that one double precision operation is
carried out practically for every
third clock cycle of the 450 MHz,
whereas the nominal maximum of the 
processor is one operation for
every second clock cycle. As it was mentioned above, without assembly 
programming an approximate reduction factor of two  
in the performance was observed. 

Increasing the volume of the simulated system one can divide
the lattice between 2 nodes (the $2\times 4\times 4$ topology
has 2 nodes in one of the directions). For even larger lattices
one can use 4 nodes (4 in one direction), 8 nodes
($2\times 4$ in two directions) 16 nodes ($4\times 4$ in
two directions) or 32 nodes ($2 \times 4 \times 4$ in three
directions). Again, the most economical way to perform the simulations
is to prepare one thermalized configuration and put it on
other nodes (this method obviously can not be used for the
$2\times 4\times 4$ topology, because in this case the whole 
machine with 32 nodes is just one lattice).

\begin{figure}[!ht]
\centerline{\includegraphics[width=8.0cm]{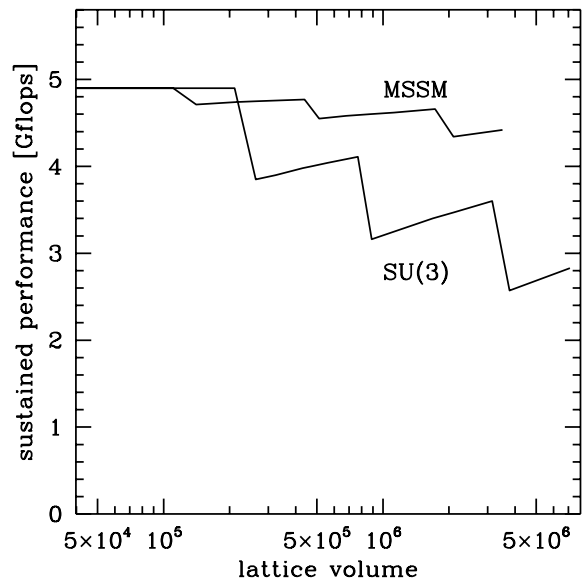}}
\caption[a]{{\sl Sustained performance of  PMS1  as a function
of the lattice volume for pure SU(3) gauge theory and for MSSM. 
The endpoints of the lines correspond to the largest
volumes which can be simulated on a 32 PC cluster.}}
\label{perform}
\end{figure}

Based on our measurements we determined the sustained performance
of a 32-node PMS cluster as a function of the lattice volume. The
result for a set of lattice volumes for finite temperature systems 
with temporal extension,
$L_t=6$ for SU(3) and $L_t=4$ for MSSM can be seen on Figure 
\ref{perform}. Clearly, the largest volume one can reach is approximately
twice as large for the SU(3) gauge theory than for the MSSM. 
For both cases there are regions where the performance increases
with the volume. This can be easily understood to happen due to the fact 
that larger volume means better surface/volume ratio, thus better performance.
There are three drops in the performance for both SU(3) and MSSM. 
They correspond to
lattice volumes for which new communication directions were opened
(or, in other words, the dimension of the mesh of the nodes on which the lattice
was divided, increased by one) in order to fit the lattice into the 
available RAMs. As it can be seen the performance for the MSSM 
is still very high even at the largest volume with three-dimensional
communication: it is just 10\% smaller than the performance without
communication. This plot gives us the optimum architecture of
such a parallel computer. The number of nodes and the
number of communication directions used for a given lattice 
should be as small as possible simultaneously. 
This means  a $2\times 4\times 4$
topology for 32 nodes and a $2\times 4\times 8$ topology 64 nodes. 

Despite the fact that the speed of the communication between two nodes
is not that high (2 Mb/s) the performance of the cluster is quite good. The
reason for this is the high speed of the individual nodes (450 MHz) and the
large RAM on each node. This sort of design does not need a division of the
lattice to hundreds of sub-lattices, thus it does not need a very fast
communication. 

The total sustained performance of PMS1 for
double precision calculations is $\approx$4Gflops. The price/(sustained
performance) ratio is quite excellent:\$3/Mflops.

For single precision simulations one can use the MMX instruction set which is
8 times faster than the double precision operations (4 operations for each
clock cycle). We estimated the single precision performance by assuming that
MMX programming results in 20\% decrease in performance. The total performance
of PMS1 is $\approx$20Gflops with 0.60\$/Mflops price-to-sustained performance
ratio. 
The PMS1 machine, similarly to other workstation farms has a moderate 
maximum sustained performance as compared to Teraflop-scale machines 
(CP-PACS \cite{CP-PACS} or QCDSP \cite{QCDSP}). 
However, PMS1 has a  much better price/(sustained performance) ratio than 
other workstation farms.

\chapter{SUMMARY} \label{sect_sum}

In my thesis  two phenomena pointing beyond the Standard Model were discussed.
In chapter \ref{MSSM} the possibility of baryogenesis in the minimal
supersymmetric extension of the Standard Model was examined. The problem
cannot be solved fully by perturbation theory so lattice simulations were
needed. My main contribution to this analysis is related to this
non-perturbative part.

A 5000 line C program was written 
that performs simulations of the MSSM. During
the writing of this program the Monte-Carlo methods used in
earlier works had to be improved. 
The overrelaxation and heatbath algorithms for the scalar fields
were introduced in section \ref{sect_monte}.

Using the MSSM program, systematic finite temperature and zero
temperature simulations were performed. First the phase transition point was
determined at finite temperature for different volumes and infinite volume
extrapolation was carried out. 
Then at the same physical point zero temperature simulations
were performed and the mass spectrum was determined. This whole procedure was
repeated for four different lattice spacings.

After observing the good agreement with perturbation theory, the lattice
results obtained at different lattice spacings can be perturbatively 
corrected to be on a line of constant physics (LCP) and continuum limit
extrapolation can be done.

For perturbation theory it is important to know the renormalization corrections
to the squark masses. To this end the phase diagram in the
$m_U^2$--$T$ plane was determined. 
The transition point to the color-breaking phase is a
good reference point for squark mass renormalization.

The produced baryon number is in strong connection with the shape of the 
bubble wall during the phase transition. Using constrained simulation, the 
profile of the bubble wall can be measured. With an appropriate shifting
procedure the wall profile and the wall width were determined. The 
variation of the $\beta$ parameter through the bubble wall was also 
found.

The other phenomenon which was discussed in chapter \ref{UHECR} is connected
to ultrahigh energy cosmic rays. The highest energy cosmic rays have 
macroscopic energy which cannot be explained with our present knowledge.
The interesting clustering features of the highest energy events lead to 
the assumption that these ultrahigh energy cosmic rays may come from compact
sources. Based on the observed clustering properties the density
of these sources was determined
for arbitrary spatial, energy and luminosity distributions. 
Three examples for energy distribution and two for luminosity distribution
were studied in detail.

A useful function of the analysis was the $P(r,E,E_c)$ 
probability function which gives the probability that a proton starting with
energy $E$ has its energy above $E_c$ after traveling a distance of $r$.
With a Monte-Carlo analysis this function was computed for a wide range of 
parameters. Thus using this parametrization the observed spectrum can be
calculated to any injected spectrum without further Monte-Carlo simulations.

The fact that some events have extremely high energies gives the 
possibility that they might not be accelerated from lower energies (these 
are the so called ''bottom-up'' scenarios), but they are the decay products
of some heavy metastable particle. These scenarios are commonly called
''top-down'' scenarios.

Based on this assumption the decay spectrum of superheavy 
particles was determined. For this the fragmentation functions of protons
at these high energies were needed. 
Since this is a non-perturbative quantity,
the experimental results for the fragmentation functions at low energy had
to be used and these were evolved to high energies. 
This fragmentation function was combined
with the prediction of the Modified Leading Logarithmic Approximation (MLLA)
which gives the small momentum region of the fragmentation functions more
accurately.

The lattice simulations of MSSM and the Monte-Carlo study of the propagation
of cosmic rays required a huge amount of CPU time. We decided to build
a cost-effective supercomputer from PC elements, the Poor Man's 
Supercomputer (PMS). In the fourth chapter of my thesis the
hardware and software architecture of this machine was described. 
The most important
hardware solution is the special communication card that provides fast 
communication between adjacent nodes, while on the software
side the job-management system and the kernel driver for the cards were
designed. 

\section{Acknowledgments}
First of all I would like to 
thank the help and patience of my advisor Zolt\'an Fodor.
He has a great part in my results with lots of ideas and encouragement.
I am also grateful to Ferenc Csikor for many useful discussions.
It was a pleasure to work together with P\'al Heged\"us, Viktor Horv\'ath, 
Antal Jakov\'ac and Attila Pir\'oth.
Finally I thank B.A Kniehl for the fragmentation function of the proton
generously supplied prior to its publication.

\end{document}